\documentclass[sigconf]{acmart}

\AtBeginDocument{%
  \providecommand\BibTeX{{%
    \normalfont B\kern-0.5em{\scshape i\kern-0.25em b}\kern-0.8em\TeX}}}

\renewcommand\footnotetextcopyrightpermission[1]{} % removes footnote with conference information in first column


\copyrightyear{2022}
\acmYear{2022}
\setcopyright{acmlicensed}\acmConference[RAID 2022]{25th International
Symposium on Research in Attacks, Intrusions and Defenses}{October 26--28,
2022}{Limassol, Cyprus}
\acmBooktitle{25th International Symposium on Research in Attacks, Intrusions
and Defenses (RAID 2022), October 26--28, 2022, Limassol, Cyprus}
\acmPrice{15.00}
\acmDOI{10.1145/3545948.3545963}
\acmISBN{978-1-4503-9704-9/22/10}

\usepackage{enumitem}
\usepackage{diagbox}
\usepackage{array,multirow,graphicx,makecell}
\usepackage{listings,amsfonts}
\usepackage{bigstrut}
\usepackage[linesnumbered,ruled,vlined]{algorithm2e}
\usepackage{booktabs}
\usepackage{bm}
\usepackage{hyperref}
\usepackage{framed}
\usepackage{booktabs,caption}
\usepackage{subcaption}
\usepackage{amsmath,xcolor,pifont}
\usepackage{xspace}
\usepackage{float}
\usepackage{color}
\usepackage{ifthen}
\usepackage[T1]{fontenc}
\usepackage{enumerate}
\usepackage{balance}
\usepackage{tikz}
\usepackage{calc}
\usepackage{url}
\usepackage{hyperref,endnotes}
\usepackage[misc]{ifsym}
\usepackage{amsmath}

\newcommand{\tab}{\hspace*{1em}}
\newcommand{\code}[1]{{\fontfamily{cmtt}\fontseries{m}\fontshape{n}\selectfont\small{#1}}}

\newcommand{\term}[1]{\( \mathcal{#1} \)\xspace}
\newcommand{\equa}[1]{\( #1 \)\xspace}
\newcommand{\dapp}{DApp\xspace }
\newcommand{\ua}{{\term{UA}}\xspace}
\newcommand{\za}{{\term{ZA}}\xspace}
\newcommand{\oa}{{\term{OA}}\xspace}
\newcommand{\RA}{{\it RiskAmount}\xspace}
\newcommand{\RL}{{\it RiskLevel}\xspace}
\newcommand{\cmark}{\ding{51}}%
\newcommand{\xmark}{\ding{55}}%

\pdfoutput=1
\begin{document}\sloppy
\title{Penny Wise and Pound Foolish: Quantifying the Risk of Unlimited Approval of ERC20 Tokens on Ethereum}

% Author Information
\author{Dabao Wang}
\affiliation{
 \institution{Monash University}
 \city{Melbourne}
 \country{Australia}
}
\email{dabao.wang@monash.edu}

\author{Hang Feng}
\affiliation{
 \institution{Zhejiang University}
 \city{Hangzhou}
 \country{China}
}
\email{feng.h@zju.edu.cn}

\author{Siwei Wu}
\affiliation{
 \institution{Zhejiang University}
 \city{Hangzhou}
 \country{China}
}
\email{wusw1020@zju.edu.cn}

\author{Yajin Zhou}
\affiliation{
 \institution{Zhejiang University}
 \city{Hangzhou}
 \country{China}
}
\email{yajin\_zhou@zju.edu.cn}

\author{Lei Wu}
\authornote{Corresponding Author: Lei Wu (lei\_wu@zju.edu.cn).}
\affiliation{
 \institution{Zhejiang University}
 \city{Hangzhou}
 \country{China}
}
\email{lei\_wu@zju.edu.cn}

\author{Xingliang Yuan}
\affiliation{
 \institution{Monash University}
 \city{Melbourne}
 \country{Australia}
}
\email{xingliang.yuan@monash.edu}

% abstract
\begin{abstract}
    The prosperity of decentralized finance motivates many investors to 
    profit via trading their crypto assets on decentralized 
	applications ({\dapp}s for short) of the Ethereum ecosystem.
    Apart from Ether (the native cryptocurrency of Ethereum), many ERC20 (a widely used token standard on Ethereum) tokens
    obtain vast market value in the ecosystem.
	Specifically, the approval mechanism is used to
	delegate the privilege of spending users' tokens to {\dapp}s. 
	By doing so, the {\dapp}s can transfer these tokens to arbitrary receivers 
	on behalf of the users.
	%Unlimited approval increases the usability of a \dapp since it reduces the required 
	%interaction between the \dapp and the user. 
	To increase the usability, \textit{unlimited approval} is commonly adopted by {\dapp}s to reduce the required interaction between them and their users.   
	However, as shown in existing security incidents, this mechanism can be 
	abused to steal users' tokens.
	
	In this paper, we present the first systematic study to quantify the risk of 
	unlimited approval of ERC20 tokens on Ethereum.
	Specifically, by evaluating existing transactions up to 31st July 2021, we find that unlimited approval is prevalent (60\%, 15.2M/25.4M) in the ecosystem, 
    and $22\%$ of users have a high risk of their approved tokens for stealing.
	After that, we investigate the security issues that are involved in interacting with the UIs
	of $22$ representative {\dapp}s and $9$ famous wallets to prepare the approval
	transactions. The result reveals the worrisome fact that all {\dapp}s request 
	unlimited approval from the front-end users and only 10\% (3/31) of UIs provide 
	explanatory information for the approval mechanism. Meanwhile,
	only 16\% (5/31) of UIs allow users to modify their approval amounts.
	Finally, we take a further step to characterize the user behavior into five modes
	and formalize the good practice, i.e., \textit{on-demand approval and timely spending}, towards securely spending approved tokens. 
    However, the evaluation result suggests that only 0.2\% of users follow the good practice to mitigate the risk.
	Our study sheds light on the risk of unlimited approval and provides 
	suggestions to secure the approval mechanism of the ERC20 tokens on Ethereum. 
    
\end{abstract}

% keywords

%%
%% This command processes the author and affiliation and title
%% information and builds the first part of the formatted document.
\maketitle

\section{Introduction}
\label{sec:introduction} 

As a blockchain-based financial system, Decentralized Finance (DeFi) has been blooming in recent years.
The prosperous development of 
DeFi brings a rapid growth of decentralized applications ({\dapp}s). 
Many investors are attracted to profit via trading their crypto 
assets on {\dapp}s of the Ethereum ecosystem. 
Apart from Ether (the native token on Ethereum), the ERC20~\footnote{ERC20 
is the most widely used token standard on Ethereum.} token standard~\cite{erc20} 
allows users to use their tokens in other arbitrary contracts.
Due to the convenience, ERC20 tokens rapidly become popular crypto assets 
that help boost the prosperity of {\dapp}s.
Specifically, the ERC20 token standard introduces an approval mechanism 
enabling users to delegate the privilege to {\dapp}s to spend users' tokens. 
In particular, users first grant the permission of their tokens to 
{\dapp}s with a transaction (i.e., \textit{approval transaction}). 
Then on behalf of users, {\dapp}s can launch another transaction to 
transfer users' approved tokens to provide the requested service.

As such, the front-end users 
need to construct at least two transactions to spend their ERC20 tokens, 
including the approval transaction and another transaction to transfer tokens. 
Specifically, the front-end users will directly interact with {\dapp}s’ and wallets’ UIs to construct these transactions.
To minimize the cost of repeatedly sending approval transactions, 
\textit{unlimited approval} is used to grant the privilege of using an unlimited 
number of users' tokens to {\dapp}s. 

Despite that unlimited approval can help users save money (gas fee) from 
repeatedly sending approval transactions, it can also be \textit{abused} 
to steal users' tokens. 
For example, malicious {\dapp}s~\cite{incident_degenmoney,incident_unicat} may elaborately trick the users into granting the token approvals, and surreptitiously transfer those approved tokens (e.g., through backdoor functions). 
Moreover, many recent incidents~\cite{incident_bancor,incident_primitive,incident_defisaver} 
reveal that hackers can steal users' approved tokens (worth over $\$180M$) by exploiting vulnerabilities laid in the legitimate  {\dapp}s. 
On the other hand, this situation is exacerbated by the strategy adopted by those popular DeFi {\dapp}s with a considerable TVL~\footnote{I.e., Total Locked Value.} (e.g., Uniswap~\cite{UniswapV2}), and crypto wallets with a large number of users (e.g., Trust Wallet~\cite{trustwallet}). Specifically, the {\dapp}s tend to use unlimited approval as the default settings on their UIs to interact with the users, while the wallets usually (if not always) keep these settings for the interactions. Furthermore, both of them do not provide the functionality for users to adjust the approval amounts.

As a result, there is an urgent need to understand the risk of unlimited approval 
and mitigate attacks against the approval mechanism. 
Although there exist several studies focusing on the improper implementation~\cite{chen2019tokenscope} and use (or even abuse)~\cite{rahimian2019resolving, gao2020tracking, chen2020traveling, xia2021trade} of the ERC20 token, to the best of our knowledge, there is still a lack of study to systematically and comprehensively reveal and measure the risk of unlimited approval in the ecosystem, including:
1) {\it to what extent unlimited approval is abused and how risky of users' approved tokens?}
2) {\it what security issues are involved in interacting with {\dapp}s and wallets to prepare approval transactions?}
3) {\it What is the current status of user behaviors and how do users achieve good practice to 
spend approved tokens towards mitigating the risks?}

In this paper, we present the first systematic study of unlimited approval 
of ERC20 tokens.
By analyzing all transactions (before 31st July 2021), we
identify $25.4M$ approval transactions. Then, we analyze the distribution 
and growth trend of unlimited approval transactions, as well as the sender, 
spender, and token participating in the approval transactions. Furthermore, 
we define {\it RiskAmount} and {\it RiskLevel} to measure the potential 
risk of approved tokens ({\bf Section~\ref{sec:transaction}}). 
To gain an in-depth understanding of the abuse of unlimited approval, we 
investigate $22$ {\dapp}s and $16$ wallets from two aspects: {\it Interpretability} and {\it Flexibility}. 
{\it Interpretability} indicates the status of explanatory information that the UIs provide for users to understand the risk of unlimited approval, 
while {\it Flexibility} indicates whether the UIs provide the modification feature for users to adjust the approval amount. 
By doing so, our investigation can disclose whether {\dapp}s and 
wallets guide users to securely construct approval transactions ({\bf Section~\ref{sec:interaction}}).
Finally, to further understand the user behavior of spending approved tokens,
we first detect users' temporal sequence of approving and spending actions,
as well as characterize user behaviors into five modes. Then, we analyze the 
distribution of the user behavior that follows the good practice of using 
unlimited approval ({\bf Section~\ref{sec:user}}).

To sum up, this study reveals a number of worrisome facts about unlimited approval:
\begin{itemize}[leftmargin=*]
    \item {\bf Unlimited approval is prevalent (60\%, 15.2M/25.4M) on Ethereum 
    and 22\% of users have a high risk of their approved tokens for stealing.} 
    We identify $25.4M$ approval transactions and 60\% of them are 
    unlimited approval. Meanwhile, 60\% of unique users participate in
    the unlimited approval transactions.
    The evaluation result of the {\it RiskAmount} and {\it RiskLevel} on the top three tokens 
    (i.e., USDC, USDT, and DAI) suggests that $22\%$ of 
    users are threatened by token stealing with a high risk.
    \item {\bf 
    Most {\dapp}s and wallets do not provide 
    comprehensive understandings (around 90\%) and flexibility (84\%) for users to mitigate the risk of unlimited approval.}
    We discover that all selected {\dapp}s ($22$) request unlimited approval from users. 
    However, only 10\% {\dapp}s and wallets (i.e., 9\% (2/22) and 11\% (1/9), respectively) provide explanatory information on the approval mechanism.
    Moreover, only 16\% (5/31) of UIs enable users to adjust the approval amounts.
    Surprisingly, we also find two {\dapp}s (i.e., Curve Finance~\cite{curve} and Yearn Finance~\cite{yearn}) mislead users to construct unlimited approval transactions.
    \item {\bf 
    Only 0.2\% of user behaviors follow the good practice to spend approved tokens. } 
    After the in-depth investigation of users' transactions regarding 
    the approval mechanism, we characterize the user behavior into five modes
    and formalize the good practice of spending approved tokens. 
    Through this, we further measure the number of user behaviors following the good practice. 
    The result suggests that 76\% of user behaviors follow modes 1 and 2, and 99\% of 
    their user behaviors are using unlimited approval.
    However, only 0.2\% (2,475/1,496,886) of detected user behaviors
    securely spend approved tokens, which suggests that most users are not aware of the risk.
    
\end{itemize}

\section{Background}
\label{sec:background}

\subsection{Ethereum Primitives}

\noindent {\bf Accounts.}\tab
Ethereum has two types of accounts: External Owned Account (EOA) and Contract 
Account (CA). EOAs are controlled by private keys owned by %\review{A4\_Minor}{\remove{external}} 
users. In
contrast, CAs are controlled by smart contracts, which are known as snippets
of JavaScript-like code. To create a CA, users need to send a signed transaction
to deploy their smart contracts on the Ethereum blockchain.

\noindent {\bf Transactions.}\tab
\label{subsec:background_txn}
Ethereum blockchain is a state machine that can be 
altered by validated transactions. An EOA can initiate a transaction based on 
different purposes such as transferring Ether, invoking a function of a smart 
contract or deploying a smart contract. However, during the execution of a 
transaction, a transaction can trigger more transactions by invoking functions 
of other smart contracts. Therefore, to distinguish them, we call transactions 
initiated by EOAs as external transactions and transactions triggered within an 
external transaction as internal transactions. In Section~\ref{sec:transaction}, 
the word `transaction' indicates the external transaction.

\noindent {\bf Smart Contract.}\tab 
A smart contract on Ethereum is an account controlled by an 
immutable program (with many executable functions) written 
in a high-level language (i.e., Solidity).
To execute the smart contract, one can send transactions by invoking
the function of the smart contract.
A function of an Ethereum smart contract can be identified by the function signature.
The function signature contains the first four bytes of the hash value (SHA3) of
the function name with the parenthesized list of parameter types.
Initiating a transaction by invoking a function of a smart contract 
requires a user to provide a function signature with corresponding parameters in the 
call data section. Then, the transaction will be executed based on the logic of the
smart contract. 

\subsection{Decentralized Applications.}
The decentralized application (\dapp) is an application running in the Ethereum
system with accessible and transparent smart contract(s). In addition, the \dapp normally 
builds a user interface (e.g., website) for front-end users to use its services. 
In the following, we introduce the high-level ideas of two popular types of {\dapp}s 
to help understand this work.

\noindent {\bf Decentralized Exchanges (DEXes).}\tab 
Unlike centralized exchanges (CEXes), decentralized exchanges (DEXes) 
are exchanges without any centralized authority. Instead, users are allowed to exchange
their cryptocurrencies with full control of their capital. Moreover, there are 
two main categories of DEXes: automated market maker and order book. 

\noindent {\bf Lending Platforms.}\tab
Lending platforms in DeFi offer loans of cryptocurrencies for users without any 
intermediary. Borrowers can directly take loans on the lending platforms by 
paying interest periodically. Moreover, most lending platforms have 
their liquidation mechanism (with different collateral ratios) to protect 
their loaned assets from serious price slippage of borrowers' collateral.

\subsection{Wallets}
In the cryptocurrency ecosystem, the users maintain a pair (or multiple pairs) of public 
and private keys to claim cryptocurrency ownership. Specifically, the users 
use public keys to receive cryptocurrencies and private keys to transfer their 
cryptocurrencies. In this context, the wallets function like a bank account to 
manage users' cryptocurrencies by storing their key pairs safely. The purpose of
a secure wallet is to prevent users' private keys from leaking. There are 
two main types of wallets: software and hardware wallets.
The {\it software} wallets are designed as software available in three main formats
(e.g., Mobile, Desktop, and Web-based Extension). Users' information (e.g., 
private keys) is regularly stored online or on local devices. 
The {\it hardware} wallets leverage some hardware devices (e.g., hard drive) to store
users' information offline.

\subsection{ Approval Mechanism of ERC20 Tokens}
\label{subsec:approval_mechanism}

To understand the approval mechanism of ERC20 tokens,
we elaborate on the details of two variables and two functions  
implemented in the ERC20 token contract regarding the approval
mechanism:

\begin{itemize}[leftmargin=*]
    \item \code{mapping(address=>uint) {\bf balanceOf}}: 
    \code{balanceOf} is a mapping list recording the amount of tokens owned by users. 
    \item \code{mapping(address=>mapping(address=>uint)) {\bf allowance}}:
    \code{allowance} is a mapping list recording the amount of tokens approved by users to spenders. 
    
    \item \code{{\bf approve}(address \_spender, uint256 \_amount)}: 
    \code{approve} is the function executed by accounts (i.e., CA or EOA) to 
    grant the permission of the tokens to indicated spenders. 
    The approval transaction will update the variable \code{allowance}.
    
    \item \code{{\bf transferFrom}(address \_sender, address \_receiver, uint256 \_amount)}:
    \code{transferFrom} is the function executed by approval recipients
    to transfer tokens. The transaction including the invocation of 
    \code{transferFrom} will update both \code{allowance} and \code{balanceOf}.
    
\end{itemize}

Apart from the native coin (Ether) on Ethereum, various ERC20 tokens 
circulate in the Ethereum ecosystem and have obtained a great market 
value (e.g., USDC~\cite{usdc} and USDT~\cite{usdt}). 
The ERC20 token standard, introduced in 2015, is one of the most 
popular token standards. 
Different from Ether, there are two ways to spend ERC20 tokens.
First, the user can directly transfer their ERC20 tokens to other 
accounts by invoking the \code{transfer} function. 
Second, according to the approval mechanism, the user can delegate their 
ERC20 tokens 
~\footnote{Named as \textbf{approved tokens} in this paper.}
to other accounts 
by invoking the \code{approve} function 
~\footnote{Named as \textbf{approval transactions} in this paper.}. 
Once the 
user's approval transaction is processed, the accounts that received approval
can spend the user's approved tokens by invoking \code{transferFrom} 
function~\footnote{Named as \textbf{execution transactions} in this paper.}.
Note that spending ERC20 tokens on {\dapp}s normally requires users to
send approval and execution transactions sequentially 
by interacting with {\dapp}s and wallets (see Section~\ref{subsec:int_cons}). 
However, if 
the user's approved tokens are used out by the {\dapp}, the user 
has to construct an approval transaction again before requesting 
services from the {\dapp}. Therefore, users often construct unlimited approval transactions to save money 
from repeatedly sending approval transactions with limited 
approval amounts.
To explore the good practice of spending approved tokens, 
the user behaviors (e.g., how do users specify the approval amounts and construct the corresponding transactions) and their distribution need to be characterized and analyzed (see Section~\ref{sec:user}).

\section{Attack Models and Motivating Examples}
\label{sec:atk}
\begin{figure}[h!]
	\centering
	\includegraphics[width=1\linewidth]{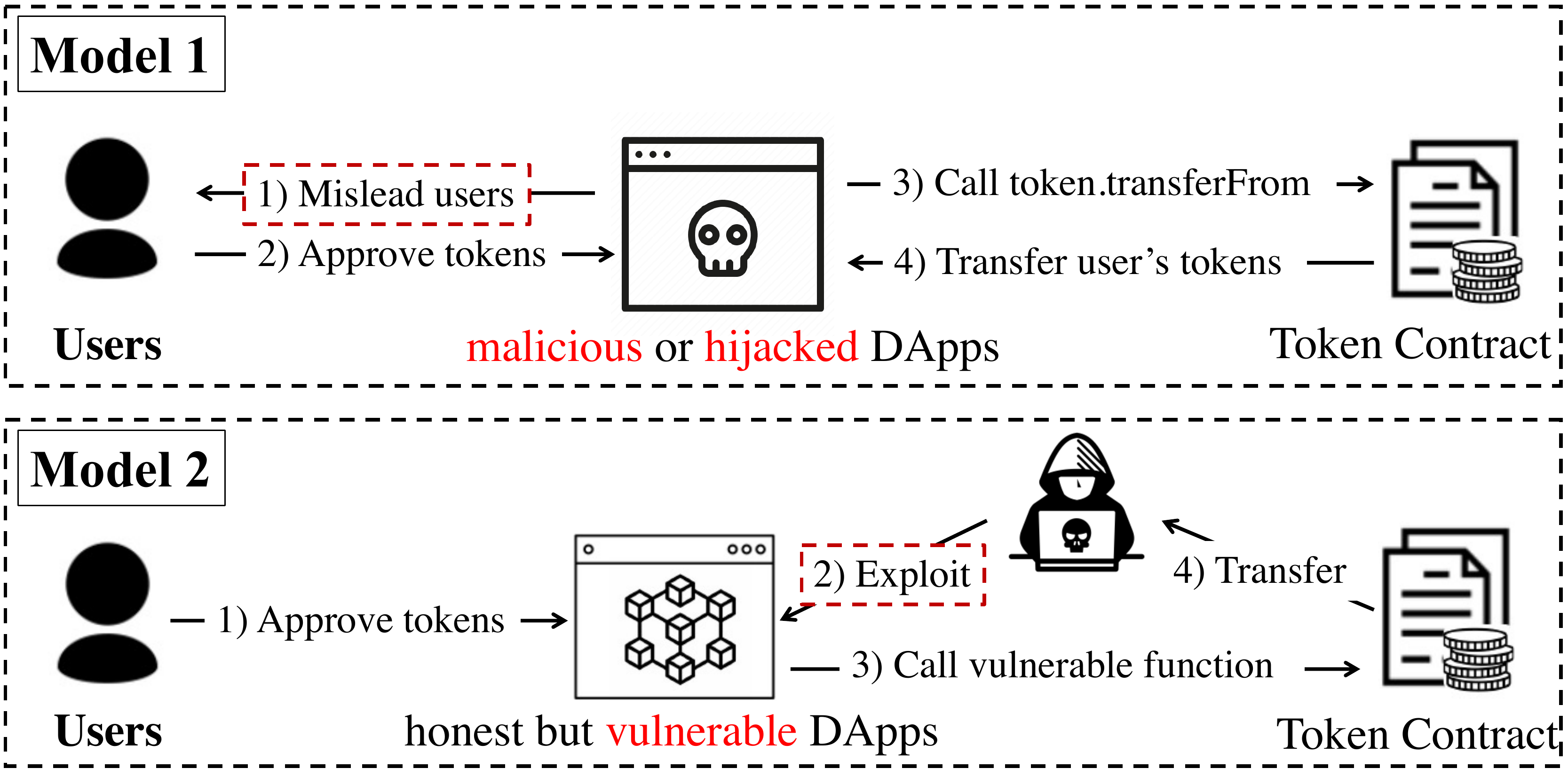} 
	\caption{\bf The attack models of stealing users' approved tokens.} 
	\label{fig:atk}
\end{figure}

In this section, we will elaborate on the high-level ideas of some infamous attacks to understand the risk of the approved ERC20 tokens.
Specifically, our investigation suggests that users' approved tokens might be stolen in two ways: 
1) approving tokens to malicious or hijacked {\dapp}s, and
2) approving tokens to vulnerable {\dapp}s.
Accordingly, the related security incidents can be classified into two attack models (as shown in Figure~\ref{fig:atk}). 
For each model, we will first describe the attack flow, then demonstrate it with a concrete example, and finally summarize the representative attacks in recent years.
Moreover, we list all incidents and summarize the loss in Table~\ref{tbl:incidents}.

\subsection{Attack Model 1} 
Attack model 1 is mainly related to the front-end of {\dapp}s.
The victims are first misled to approve their tokens 
to the malicious or hijacked {\dapp}. 
Then, with users' token approvals, the malicious or hijacked {\dapp} can 
further invoke \code{transferFrom} of ERC20 token contracts to 
steal users' approved tokens.

\noindent \textbf{A Concrete Example:}\tab
In September 2020, a {\dapp} called UniCat~\cite{incident_unicat}
stole over $\$140,000$ worthy of UNI~\cite{uni} tokens from users.
Initially, UniCat pretended as a yield farming {\dapp} promising to 
distribute great profits based on users' deposited UNI tokens.
With attractive advertisements, UniCat induced many users to approve UNI tokens 
to their accounts.
% smart contract address. 
%
After receiving token approvals from many users, UniCat triggered their 
backdoor function \code{setGovernance} to steal all users' tokens.
Most importantly, even if the user did not deposit all his/her UNI tokens in UniCat, 
with unlimited approval, UniCat can still steal all of the user's UNI tokens by internally 
triggering \code{transferFrom} with the backdoor function.

\noindent \textbf{Other Representative Attacks:}\tab
On 28th August 2020, Nik Kunkel~\cite{incident_degenmoney} revealed a double 
approval exploit launched by the project called Degen Money. The approach used
by Degen Money is more straightforward than what was done in the Unicat 
incident. Concretely, the UI provided by Degen Money intentionally 
tricked users to send two approval transactions to a functioning contract and
a different address, where the latter stole users' tokens. On 
2nd December 2021, the front-end of Badger was manipulated by a
malicious party~\cite{incident_badger}.
As a result, users who interacted with Badger's UI automatically sent approval 
transactions to delegate their tokens to the injected addresses, which led to a loss of $120M$ USD.

\subsection{Attack Model 2} 
In attack model 2, users lost their approved tokens by sending token approvals 
to an honest but vulnerable {\dapp}. Specifically, the attacker first detected 
the vulnerability of the {\dapp}'s smart contracts (i.e., lack of access control).
Then, the attacker internally triggered the \code{transferFrom} function using {\dapp}'s 
identity to transfer users' approved tokens.

\noindent \textbf{A Concrete Example:}\tab
On 27th Feb 2021, a {\dapp} called Furucombo experienced a 
serious hack~\cite{incident_furucombo}. The root cause is that Furucombo forgot to initialize the implementation contract address while 
immigrating the flash loan~\cite{qin2020attacking,wang2020towards} service from 
AAVE~\footnote{AAVE is the first {\dapp} officially announcing the flash loan service.}. 
Therefore, a hacker had the chance to take over the place and injected his 
malicious contract to gain access control of users' approved assets. As a result, 
the hacker stole more than $\$15M$ worthy tokens that had been approved to Furucombo
by users. 

\noindent \textbf{Other Representative Attacks:}\tab
On 22nd Feb 2020, Primitive Finance~\cite{incident_primitive}, which is 
an option market protocol, launched a set of white-hat attacks to preserve 
over $\$10M$ worthy tokens approved by users. 
The high-level idea of this incident is that the malicious 
attackers can craft fake and worthless tokens to bypass the assertion in
function \code{flashMintShortOptionsThenSwap} and transfer the fund by 
leveraging the flash swap service of Uniswap.
On 19th June 2020, Bancor ran a white-hat attack on their contract to transfer 
users' approved tokens to other safe contracts. 
The root cause is that Bancor mistakenly set the function \code{safeTransferFrom}
as public, so that any user can invoke this function to transfer tokens approved to Bancor. 
As a result, $\$455,349$ worth of tokens were saved. 
Somehow, the massive transfer action was detected by some arbitrage bots to 
take an advantage of $\$135,299$. 
On 20th June 2020, a vulnerability~\cite{incident_defisaver} was disclosed to 
the DeFi Saver team. Instantly, the DeFi Saver team performed a white-hat attack 
to transfer all vulnerable funds to a safer address where they can only be 
accessed by the asset owners to withdraw their funds. However, once again, the flaw discovered in the function 
\code{SwapTokenToToken} and \code{takeOrder} allowed hackers to circumvent the 
requirements and perform their self-designed logic (e.g., transferring approved 
tokens in the Exchange protocol). 
On 19th January 2022, the platform Multichain~\cite{incident_multichain} was 
exploited by a group of hackers and lost around $\$44M$ because that their 
smart contract allows users to invoke \code{permit} function (see more
details in Section\ref{subsec:dis_solu}) for ERC20 tokens, which do not adopt 
\code{permit} function for the approval mechanism.

\begin{table}[t]
	\centering
	\caption{\bf The loss summary of 8 real-world incidents.}
	\label{tbl:incidents}
	\centering
	\begin{center}
	\begin{tabular}{cp{1.5cm}<{\centering}cp{1.5cm}<{\centering}}
		\toprule      	
		{\bf Attack Model} & {\bf Date} & {\bf Platform}  & {\bf Loss}   \\
		\toprule
		\multirow{3}{*}{Model 1} &$2020.08.28$& {\it Degen Money}   & -\\
		&$2020.10.05$ &{\it Unicat}& $\$0.14M$\\
		&$2021.12.02$ &{\it Badger}& $\$120M$\\
		\midrule		
		\multirow{5}{*}{Model 2} &$2020.02.22$&{\it Primitive Finance}& $\$10M$*\\
		& $2020.06.19$ & {\it Bancor} & $\$0.14M$* \\
		&$2020.06.20$&{\it DeFi Saver}    & $\$3.5M$*  \\
		&$2021.02.27$&{\it Furucombo}  	& $\$15M$ \\
		&$2022.01.19$&{\it Multichain}  	& $\$44.5M$ \\
		\bottomrule
		\end{tabular}
		\newline
	The value with `*' represents the fund which is eventually saved. 
	\end{center}
	\vspace{-15pt}
\end{table}

\section{Research Design}
\label{sec:method}

\subsection{Terminology} 
\label{subsec:term}
To favor the readability, we summarize a list of terminologies used in the rest
of the paper.
\begin{description}[leftmargin=10pt]
    \item[Approval Sender (\term{U}):] The approval sender is an Ethereum account 
    (i.e., EOA or CA) creating approval transactions to delegate permission of 
    spending their ERC20 tokens. In this study, we mainly consider the case that
    an approval sender acts as an EOA to construct approval transactions. 
    The word `user' and `sender' both indicate one EOA in the rest of this paper.

    \item[Approval Spender (\term{S}):] The approval spender is an Ethereum 
    account (i.e., EOA or CA) receiving the permission of spending approved tokens. 
    In this paper, we mainly target {\dapp}s (normally act as a CA) which spend 
    users' approved tokens based on the logic of their smart contracts.

    \item[ERC20 Token (\term{T}):] ERC20 tokens are deployed smart contracts 
    which are responsible to perform users' requests (i.e., approving and 
    transferring), and recording the correlated information (e.g., 
    \term{T}.balanceOf, \term{T}.allowance).

    \item[Wallet (\term{W}):] The wallet is normally used by front-end users
    to manage their accounts and digital assets. It also plays
    an important role to  
    help users connect to {\dapp}s and construct transactions and send it to 
    the blockchain network for confirmation.

    \item[Approval Transaction (\term{A}):] A transaction that is initiated by the ERC20 standard function
    \code{approve}. Constructing an approval transaction requires the specification
    of \term{U}, \term{S}, \term{T}, and the approval amount. Based on the 
    approval amount, we can further categorize approval transactions into three
    types: \textit{1)} Unlimited approval (\ua) indicates that the 
    approval amount of \term{A} reaches the maximum value (i.e., \code{uint256 - 1}) or the total 
    supply of token \term{T}; \textit{2)} Zero approval (\za) is 
    the approval transaction with zero approval amount. \za indicates that
    \term{U} tries to revoke their permission of their approved tokens; \textit{3)} 
    Other Approval (\oa) represents the rest of approval transactions. 

    \item[Execution Transaction (\term{E}):] A transaction contains an internal transaction 
    initiated by the ERC20 standard function \code{transferFrom}. Through invoking the function 
    \code{transferFrom}, the spender spends users' approved token based on the 
    logic of its smart contract code. The validated execution transaction will 
    further decrease the value of \term{T}.\code{balanceOf[U]} and 
    \term{T}.\code{allowance[U][S]} based on the spending amount.
\end{description}

\vspace{-10pt}
\subsection{Research Questions}
Our objective of this work is to first reveal the usage of unlimited approval in the 
ecosystem, and then provide a comprehensive empirical study towards a better understanding of the risk of abusing unlimited approval.
To this end, this study is driven by answering the following three research questions.

\noindent {\bf RQ 1:}
\textbf{What is the usage of unlimited approval in the ecosystem 
and to what extent are the potential risks taken by users?} 
Although the usage and risk of unlimited approval have been loosely mentioned in the public (e.g., the social media), the scale of unlimited approval remains unknown and there is a lack of quantification analysis for the risk of approved tokens. 
We conduct a transaction-based analysis in Section~\ref{sec:transaction} to answer this question.

\noindent {\bf RQ 2:}
\textbf{What security issues are involved in interacting with {\dapp}s and wallets to prepare approval transactions?}
Front-end users construct approval transactions through directly interacting 
UIs provided by both {\dapp}s and wallets. It is critical to understand
how the UIs of {\dapp}s and wallets guide users to construct their approval 
transactions. 
We conduct an interaction-oriented investigation 
in Section~\ref{sec:interaction} to answer this question. 

\noindent {\bf RQ 3:}
\textbf{What is the current status of user behaviors and how do users achieve good practice to 
spend approved tokens towards mitigating the risks?}
Analyzing the user behavior regarding the approval mechanism may help determine the good practice
for users to mitigate the risks.
By doing so, we provide guidelines for users to securely use approval transactions.
We conduct a user behavior analysis in 
Section~\ref{sec:user} to answer this question.

\section{Transaction-Based Analysis}
\label{sec:transaction}
In this section, we conduct the transaction-based analysis to answer RQ 1.
Specifically,
based on the previous definition of the approval transaction,
we first apply a fully-automatic approach to detect approval transactions on Ethereum (Section~\ref{subsec:tx_detect}). 
After that, we conduct a comprehensive measurement for 
identified approval transactions to gain a more detailed view of the approval's usage in the ecosystem (Section~\ref{subsec:tx_stat}). 
Finally, we further reveal the risk of users' approved tokens in terms of \RA and \RL (Section~\ref{subsec:tx_risk}).
To engage the community, we will release the collected dataset and the source code to perform the analysis on~\url{https://github.com/PanicWoo/RAID2022_Approval}.

\subsection{Approval Transaction Detection}
\label{subsec:tx_detect}

Figure~\ref{fig:ident} depicts the process of detecting approval transactions
in two stages, as follows:

\begin{figure}[t]
	\centering
	\includegraphics[width=1\linewidth]{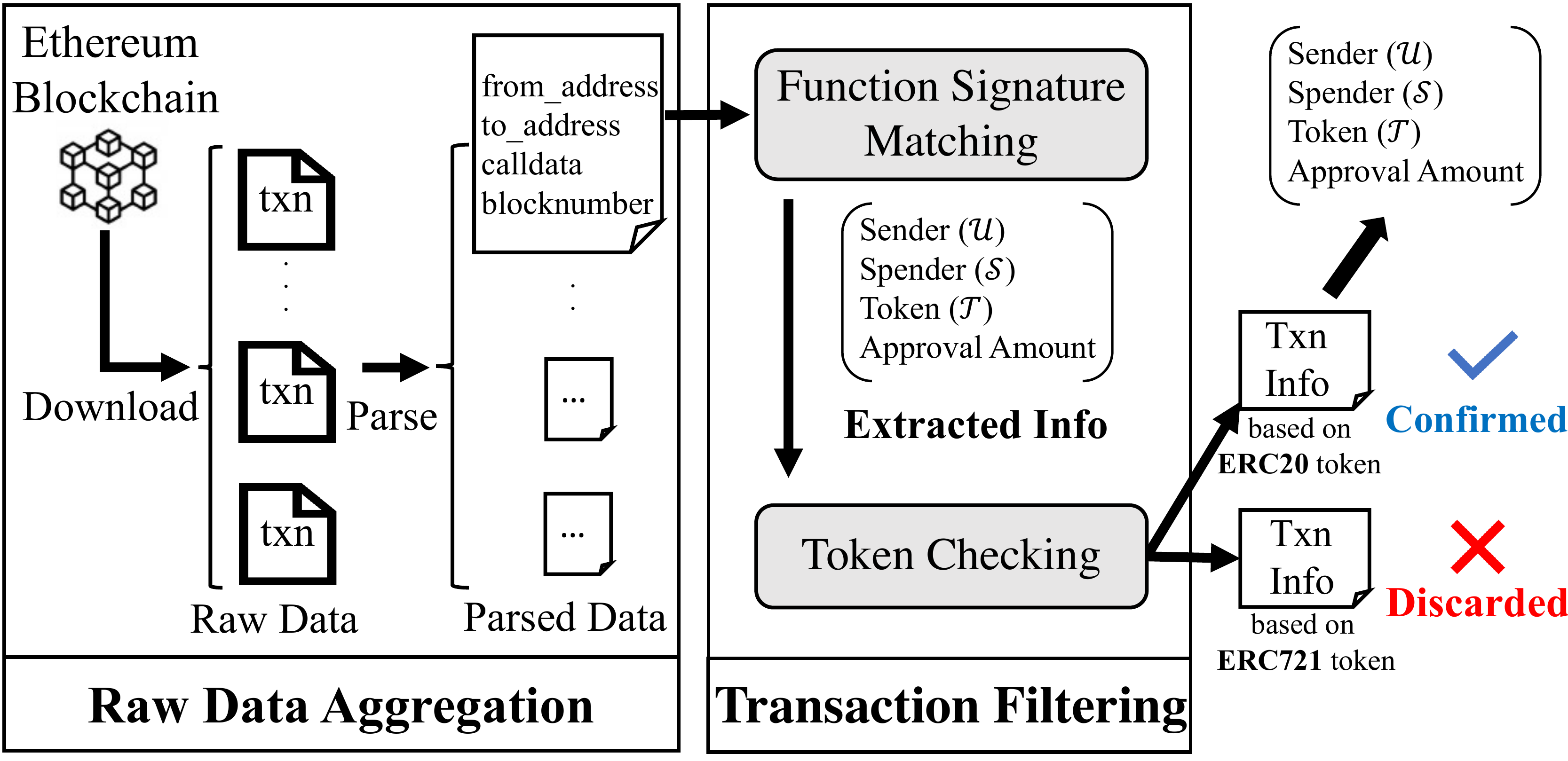} 
	\caption{\bf The process of detecting approval transactions.}
	\label{fig:ident}
\end{figure}

\noindent\textbf{Raw Data Aggregation.}\tab  
In this stage, we first take advantage of Geth~\cite{geth},
which is a popular Ethereum client, to synchronize the Ethereum ledger up to  
block $12936339$ (the latest block on 31st July 2021). 
The synchronized database contains the validated transactions and is built based on a cluster with 12 machines. Each machine consists of an Intel CPU (i7-8700 \@ 3.2GHz) and 32GB of memory.
Then, we take one more step to parse downloaded 
transactions and extract desired information (i.e., from\_address, to\_address, 
calldata, and block number) for the next stage of filtering.

\noindent\textbf{Transaction Filtering.}\tab  
In this stage, we accurately identify approval transactions through sequentially processing the parsed data with the following two components:
\begin{itemize}
    \item \textbf{Function Signature Matching. }
As aforementioned in Section~\ref{subsec:background_txn}, the call data extracted
from each raw transaction will include the invoked function signature and associated
values mentioned in the invoked function. In this component, we take the advantage of 
function signatures (Section~\ref{subsec:approval_mechanism}) provided by the ERC20 token 
standard to filter approval transactions. Once the function signature of 
the transaction is matched with the hash value of the \code{approve} function, we 
further identify the value of {\it sender}, {\it spender}, {\it token}, and 
{\it approval amount} and pass the value set to the next component. In this study, 
we only focus on the external transaction initiated by invoking the \code{approve} 
function.
\item \textbf{Token Checking. }
%{\it 2) Token Checking}: 
The \code{approve} function of ERC20 and ERC721~\cite{erc721} token standards 
have the same format and hash value. Therefore, we leverage the extracted token 
address to conduct a token checking process to reduce the false-positive rate 
of our identification process. As mentioned in the standard, ERC721 
token contracts must include the variable \code{uri}. Therefore, in this component, 
we label tokens having the variable \code{uri} in their contracts and discard 
approval transactions interacting with ERC721 token contracts.
\end{itemize}

\subsection{Overall Statistics of Approval Transactions}
\label{subsec:tx_stat}

\begin{table*}[ht]
	\caption{{\bf The overall statistic of identified approval transactions.} 
	In total, $4,935,215$ unique \term{U}, $192,172$ unique \term{S} and $94,748$ 
	unique \term{T} are identified among all collected approval transactions.}
	\label{tbl:general_sta} 
	\centering
    \resizebox{0.9\textwidth}{!}{
    \begin{tabular}{ccccc}
		\toprule      	
		{\bf Approval Type} & {\bf \# of Transaction} & {\bf Unique Sender (\term{U})} & {\bf Unique Spender (\term{S})} & {\bf Unique Token (\term{T})} \\
		\midrule
		{\bf \ua} & $15.2$M (59.8\%)& $2975 K$($60.3\%$)& $106.8 K$($55.6\%$) &$72.4 K$($76.5\%$)\\
		{\bf \za} & $0.45$M (1.8\%)& $116.1K$($2.4\%$) & $18.7 K$($9.7\%$) & $12.5 K$($13.2\%$)\\
		{\bf \oa} & $9.75$M (38.4\%)& $2777$K ($56.3\%$) & $97.5 K$($50.7\%$)& $45.5 K$($48.0\%$)\\
		\midrule
		{\bf Total Count} & ${\bf 25.4M}$ & ${\bf 4,935,215}$ & ${\bf 192,172}$ &  ${\bf 94,748}$\\
		\bottomrule
 	\end{tabular}}

		In the column of {\bf Sender (\term{U})}, {\bf Spender (\term{S})}, 
		and {\bf Token (\term{T})}, the sum of the percentages is not equal to $100\%$. 
		For instance, each sender can participate in both unlimited and zero approvals. Similarly, the \term{S} and \term{T} can be involved in different 
		types of approval transactions.
\end{table*}

In total, we discover $25.4M$ approval transactions up to block $12936339$.
Based on our definition of unlimited approval, we identify two-thirds of
approval transactions ($15.2M$, $60\%$) are unlimited approval, while $0.45M$ ($1.8\%$) and $9.75M$ ($38.4\%$) are \za and \oa respectively (the first column of Table~\ref{tbl:general_sta}). 
Note that the evaluation result is conservative due to the strict definition of {\ua}~\footnote{As a matter of fact, we also discover 23\% of approval transactions with an extremely large approval amount (greater than $2^{248}$), which could be regarded as {\ua} to some extent, as the corresponding users are still at a risk. However, we decide to follow the strict criterion to simplify the measurement.}.
Moreover, in the rest columns of Table~\ref{tbl:general_sta}, we 
discover that a great percentage of \term{U} ($60.3\%$), \term{S} ($55.6\%$), 
and \term{T} ($76.5\%$) have participated in \ua.
The statistical result indicates that \ua has been widely used in the 
ecosystem up to block $12936339$.

\begin{figure}[h]
	\vspace{-15pt}
	\centering
	\includegraphics[width=1\linewidth]{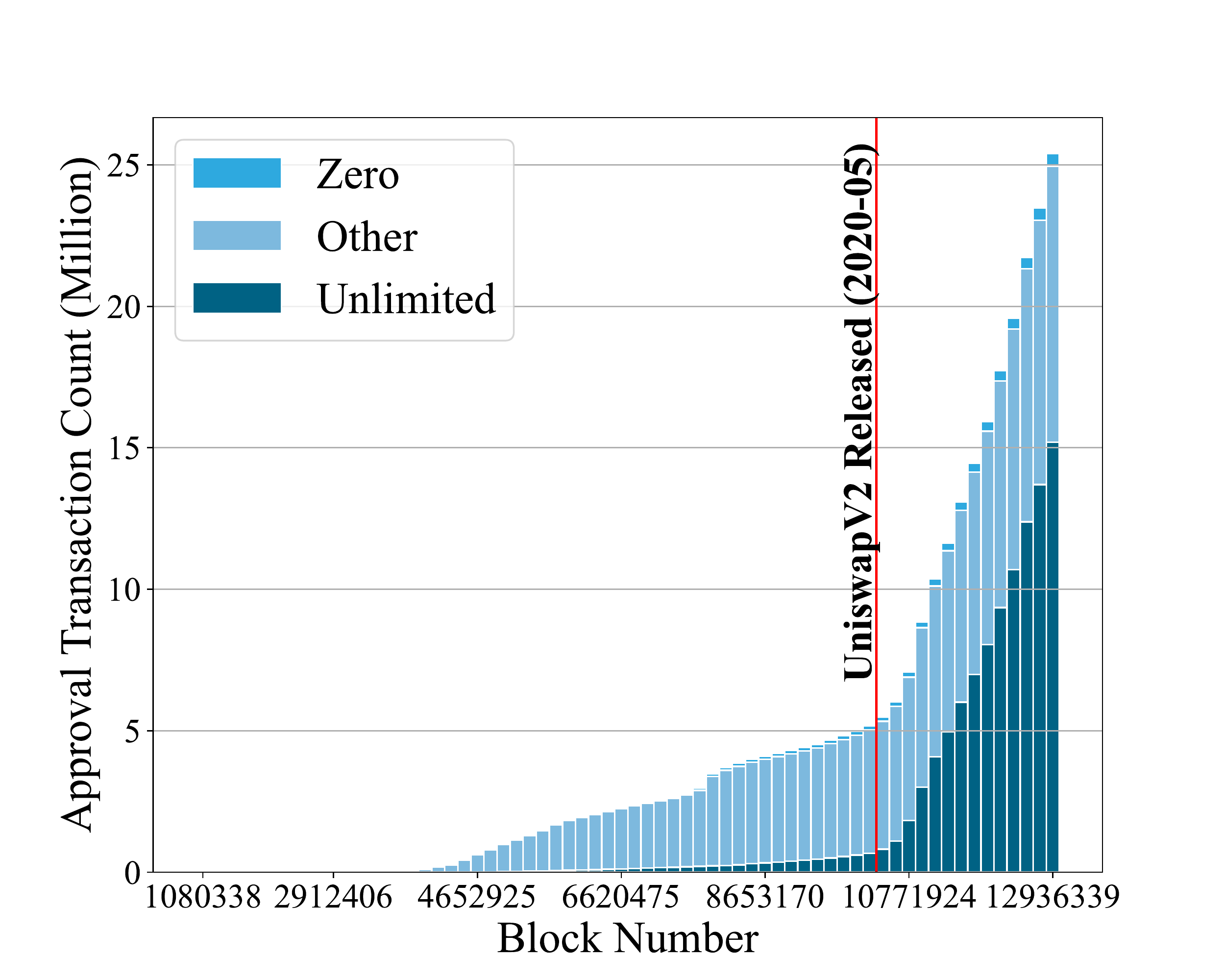} 
	\caption{\bf The growth trend and distribution of approval transactions.} 
	\label{fig:trend}
	\vspace{-10pt}
\end{figure}
\begin{figure*}[ht]
	\centering
	\includegraphics[width=1\linewidth]{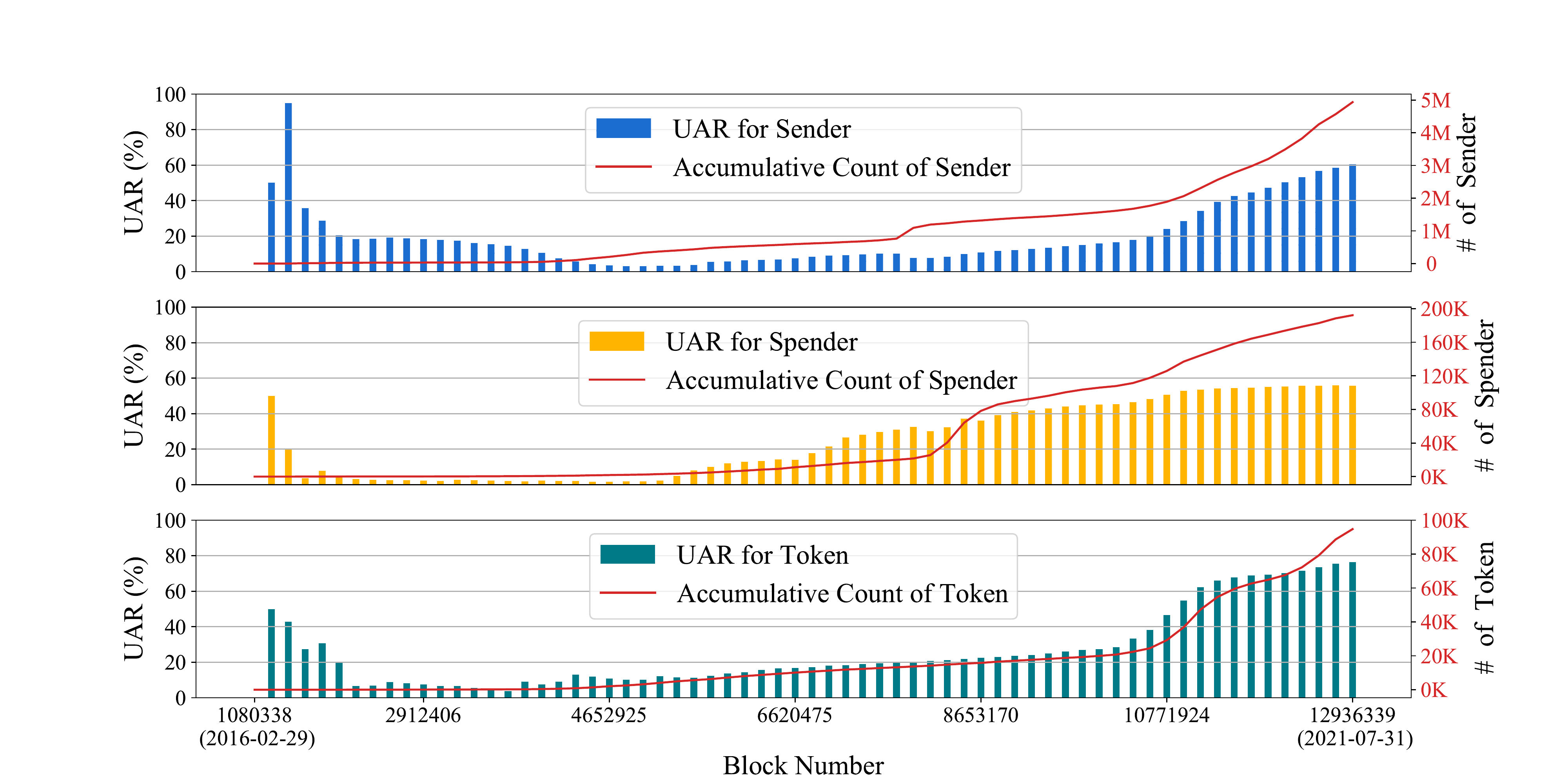} 
		The UAR of \term{U}, \term{S} and \term{T} is high
        in the early stage. The reason causing this bias is due to the small 
        number of approval transactions. 
	\caption{{\bf The growth trend of approval transaction participants (\term{U}, 
    \term{S}, \term{T}).} From the top to the bottom sub-chart, the 
    line plot presents the cumulative count of unique \term{U}, \term{S}, and 
    \term{T} respectively. The value of the line plot can be referred to the y-axis 
    on the right side. As for the bar plot, it indicates the percentages 
    of \term{U}, \term{S}, and \term{T} participating \ua (UAR in short).
    For example, the last bar of the top sub-chart shows that around 60\% of senders 
    construct unlimited approval transactions at least once. 
    The value of the bar plot can be referred to the y-axis on the left side.}
    \label{fig:trend_entities}
    \vspace{-10pt}
\end{figure*}

We also present the growth trend of approval transactions and different participants 
(\term{U}, \term{S}, \term{T}) in Figures~\ref{fig:trend} and ~\ref{fig:trend_entities}.
In Figure~\ref{fig:trend}, it is worth noting that the number of approval 
transactions has experienced a rapid increase since May 2020. In the meanwhile, 
as shown in Figure~\ref{fig:trend_entities}, the number of \term{U}, \term{S}, 
and \term{T} has also increased quickly.
The reason for the increase is the introduction of 
{\it Uniswap V2}~\cite{UniswapV2}, 
which is one of the largest DEXes on Ethereum.
Through investigating all unlimited approval 
transactions in terms of \term{S}, we find that there is $35\%$ (8.9M/25.4M) of 
approval transactions specifying {\it UniswapV2 Router} as the spender. Moreover, 
$94\%$ (8,361,001/8,925,110) of those approval transactions belong to \ua.

\subsection{Risk Analysis for Approved Tokens}
\label{subsec:tx_risk}
Unlimited approval does not directly cause users to lose their approved tokens. However, abusing unlimited approval will expose these tokens in a dangerous place. 
In the following, we will investigate the top 
three most frequently used tokens (i.e., USDC, USDT and DAI) to evaluate the potential risk of users' approved tokens. 

\begin{figure}[!htbp]
    \centering
    \includegraphics[width=0.9\linewidth]{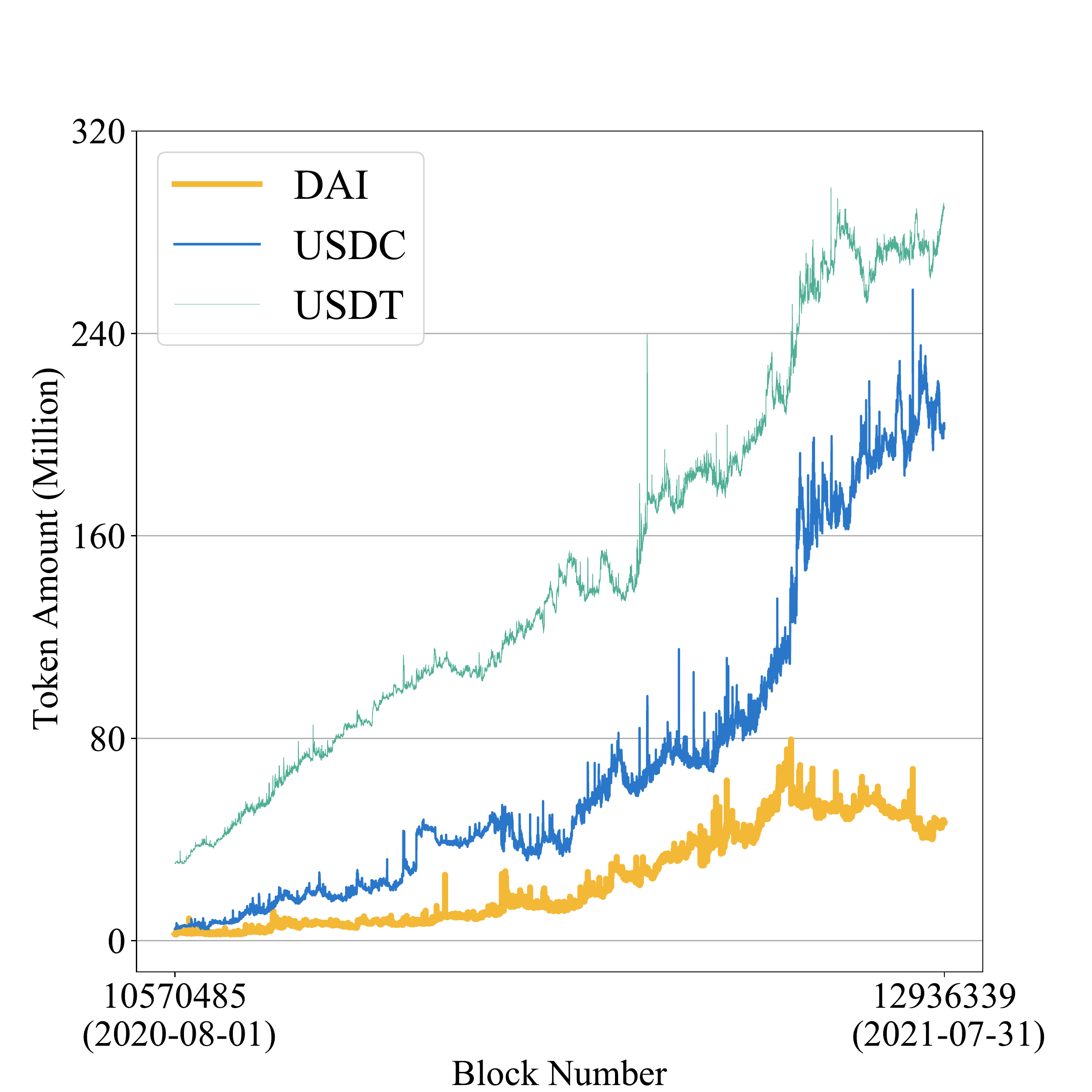}
    \caption{\bf The growth trend of risk amount on DAI, USDC, and USDT.} 
    \label{fig:ra_token}
\end{figure}
\begin{figure}[!htbp]
    \centering
    \includegraphics[width=0.9\linewidth]{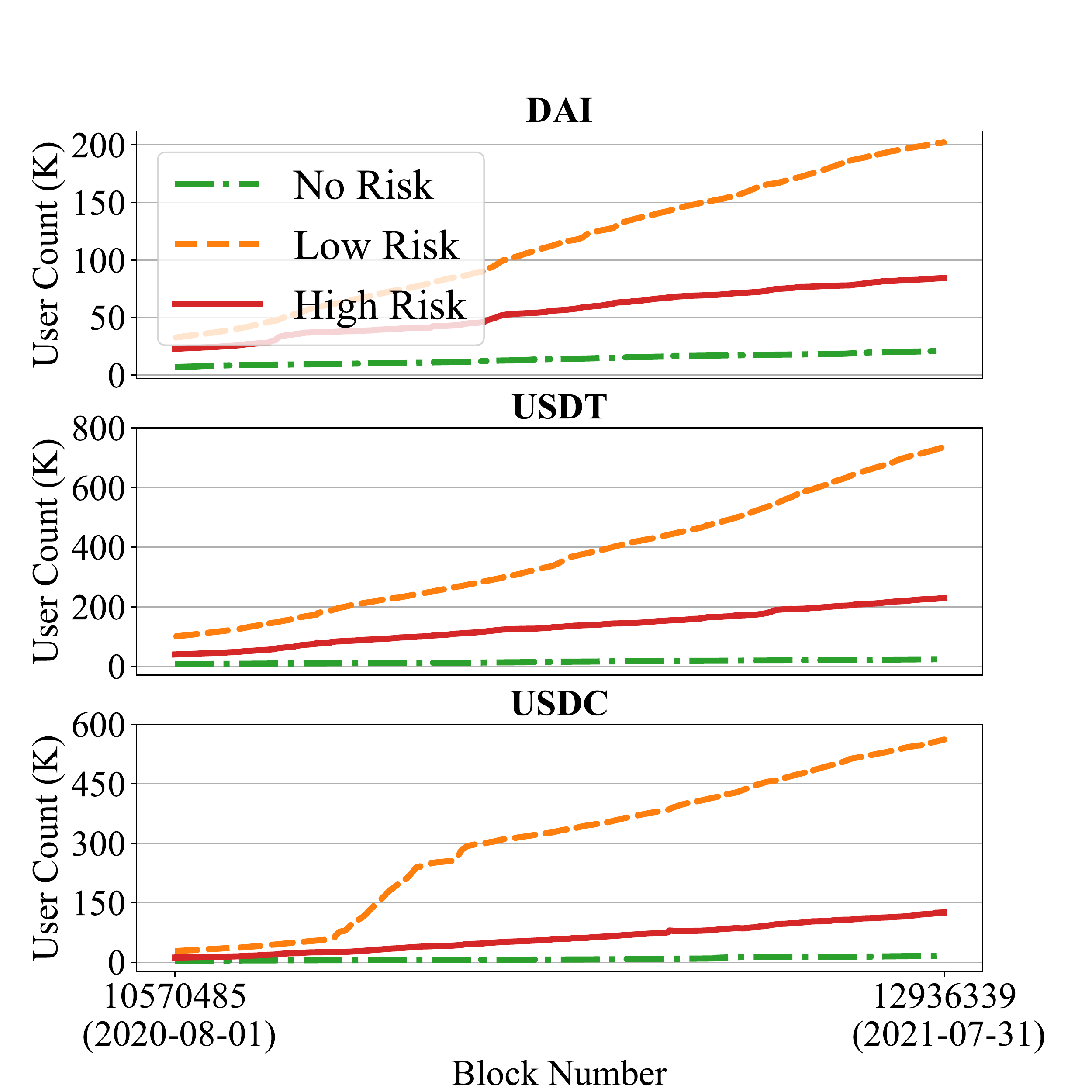}  
    \caption{\bf The growth trend of users with the different risk levels.} 
    \label{fig:user_distribution}
\end{figure}

\noindent \textbf{Risk Amount Analysis.}\tab
We define the \RA to quantify the potential risk 
of users' approved tokens. For a given tuple 
(\term{U}, \term{S}, \term{T}), the \RA indicates the amount of 
\term{U}'s owned tokens that a \term{S} can transfer. Specifically, \RA
equals the smaller amount of a \term{U}'s owned tokens and the tokens 
approved to a \term{S} by the \term{U}, as shown in Equation~\ref{eq:ra}:
\begin{equation}
\label{eq:ra}
    \hspace{-5pt}\RA= Min(\mathcal{T}.allowance{[\mathcal{U}][\mathcal{S}]},\ \mathcal{T}.balanceOf{[\mathcal{U}]})
\end{equation}
Accordingly, Figure~\ref{fig:ra_token} presents the growth trend of the accumulative \RA of all 
users in different tokens.
As for the time range, we obtain the data in the past whole year (2020-08-01 to 2021-7-31)
to plot the trend graph of \RA. 
From Figure~\ref{fig:ra_token}, we discover that \RA of each top token is generally increasing in the past whole year (especially USDT),
which suggests that more and more approved tokens are in danger as time goes by.

\noindent \textbf{Risk Level Analysis.}\tab
We define the term \RL to facilitate classifying users' approved 
tokens based on the amount of \term{U}'s owned tokens 
(\term{T}.\code{balanceOf}) and approved tokens (\term{T}.\code{allowance}). Specifically, there are three risk levels, i.e., {\it No Risk}, {\it Low Risk}, and {\it High Risk}. 
To better understand \RL, we consider a given tuple (\term{U}, \term{S}, \term{T}), and there will be Equation~\ref{eq:rl}:
\begin{equation}
    \label{eq:rl}
    \centering
    \RL =  \begin{cases}
            No\ Risk, & \mathcal{T}.allowance[\mathcal{U}][\mathcal{S}] = 0 \\
            \\
            \multirow{2}{*}{Low\ Risk,}& \mathcal{T}.allowance[\mathcal{U}][\mathcal{S}] > 0\\
             & \&\&\ \mathcal{T}.balanceOf[\mathcal{U}] = 0 \\
            \\
            \multirow{2}{*}{High\ Risk,} & \mathcal{T}.allowance[\mathcal{U}][\mathcal{S}] > 0\\ 
             & \&\&\ \mathcal{T}.balanceOf[\mathcal{U}] > 0
        \end{cases}
\end{equation}

Specifically, there are three cases: 1) if \term{U} have zero \code{allowance} to \term{S}, we consider that there is {\bf no risk} on \term{U}'s token \term{T}; and
2) if \term{U}'s have the \code{allowance} which is greater than zero and zero \code{balanceOf}, \term{U}'s token \term{T} has {\bf low risk}. Moreover, low-risk indicates that once the users receive tokens (\code{balanceOf > 0}), the malicious platforms with users' approval can instantly transfer users' received tokens; and 
3) if the \code{balanceOf} and \code{allowance} of users' tokens are both greater than zero, users' tokens are at a high-risk level and extremely risky from stealing. In the real world, many security incidents~\cite{incident_unicat, incident_bancor} demonstrate the possibility of stealing high-risk tokens.

By analyzing \RL of all users of USDC, USDT, and DAI, we observe that in the 
past year, the number of users with {\it low-risk} or {\it high-risk} tokens has grown
steadily compared to users with {\it no-risk} tokens (Figure~\ref{fig:user_distribution}).
Users with low-risk tokens become the majority among all evaluated users. 
This indicates that most users do not instantly revoke their approval from {\dapp}s. 
According to the fact, users approving tokens to malicious {\dapp}s can suffer from
token stealing once they receive corresponding tokens. 
Moreover, we further present the distribution of users with different \RL in Table~\ref{tbl:risk_analysis}. The results suggest that $97\%$ of users are underlying 
victims to token stealing, as well as $22.4\%$ of users remain their approved tokens 
with high risk. 

\begin{table}[h!]
    % \caption{The Distribution of Users in Different Risk Levels.}
    \caption{\bf The distribution of users based on risk levels.}
    % by block {\bf 12936339}.}
 	\label{tbl:risk_analysis}
	\centering
% 	\resizebox{1\textwidth}{!}{
    \begin{tabular}{ccccc}
		\toprule
        \multirow{2}{*}{\bf Token } & \multirow{2}{*}{\bf \# of Users} & \multicolumn{3}{c}{\bf Risk Level Distribution}\\
		 &  &  {\it No}  & {\it Low} & {\it High}\\
        \midrule
        {\it USDC} & $734,307$  & $2.3\%$ & $79.2\%$ & $18.5\%$ \\
		{\it USDT} & $1,058,617$  & $2.4\%$ & $74.6\%$ & $23.0\%$ \\
		{\it DAI}  & $319,913$  & $6.9\%$ & $66.0\%$ & $27.1\%$ \\
		\bottomrule
 	\end{tabular}
\end{table}

\begin{framed}
    \noindent \textbf{Answer to RQ 1:} \textit{
    Unlimited approval is widely used (60\% of all approval transactions) in the Ethereum ecosystem and a great percentage of unique users (60\%, 2.9M/4.9M) have participated in the unlimited approval transactions.
    Specifically, the number of unlimited approval transactions is extremely concentrated in {\it Uniswap V2}, which involves 35\% of approval transactions and 94\% of them are unlimited approval. 
    For the top three tokens (i.e., USDT, USDC and DAI), most users ($97\%$) are threatened by token stealing and $22\%$ of them are at high-level risk.}
\end{framed}

\section{Interaction-Oriented Analysis}
\label{sec:interaction}

Our prior transaction-based analysis suggests the prevalence of 
unlimited approval on Ethereum.
In this section, to fully understand the abuse of 
unlimited approval, we aim to reveal the security issues involved 
in interacting with the UIs of {\dapp}s and wallets by conducting an interaction-oriented investigation. 
Specifically, we first demonstrate the process of constructing 
an approval transaction from the front-end user's perspective 
(Section~\ref{subsec:int_cons}).
Then, we investigate the \textit{Interpretability} and 
\textit{Flexibility} of UIs provided by $22$ {\dapp}s 
(Section~\ref{subsec:int_dapp}) and $14$ wallets (Section~\ref{subsec:int_wallet}), respectively.

\subsection{Approval Transaction Construction}
\label{subsec:int_cons}

\begin{figure}[t]
	\centering
	\includegraphics[width=1.0\linewidth]{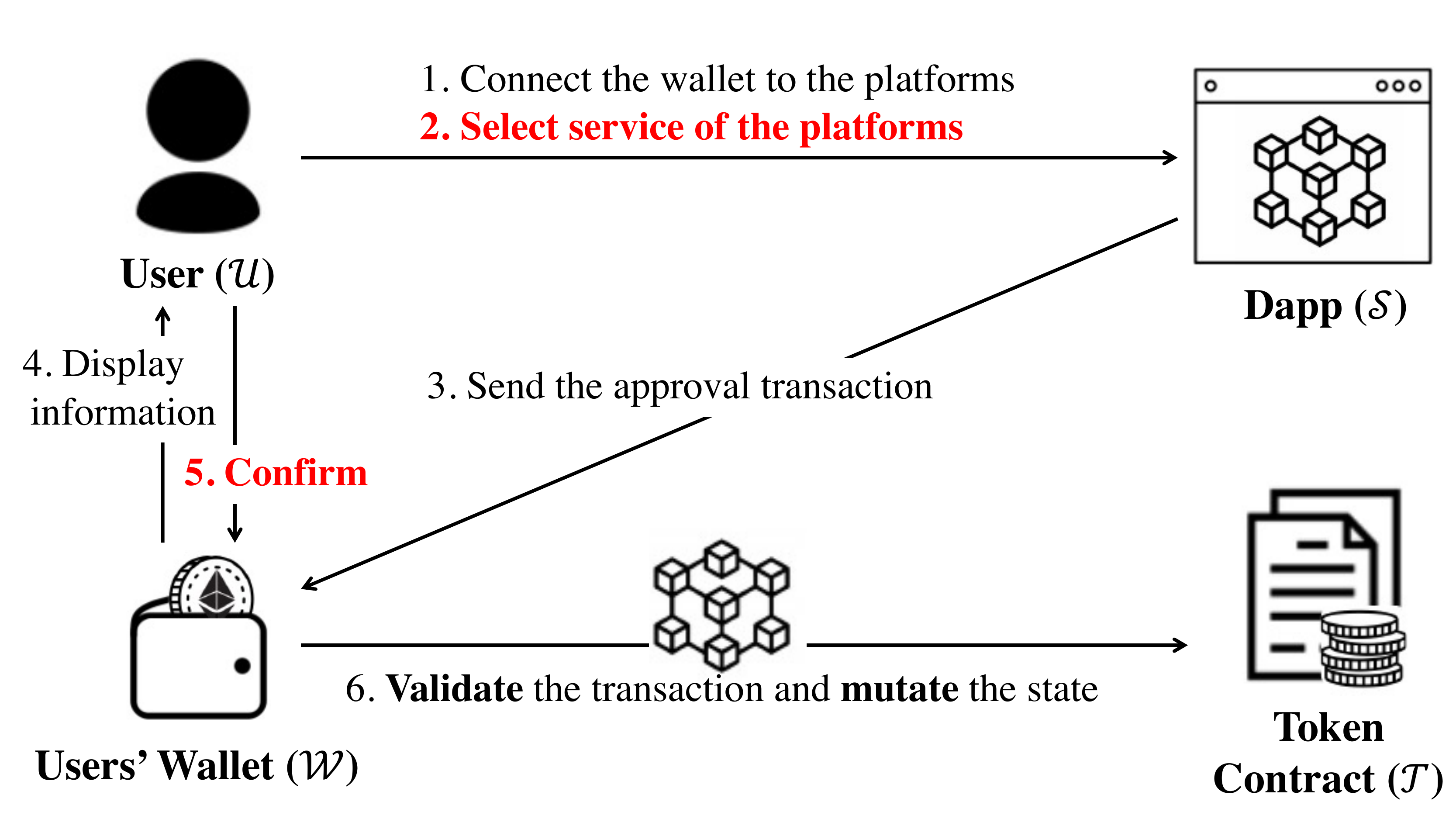}
	\caption{\bf The workflow of constructing an approval transaction by front-end users.}
	\label{fig:wkflw}
	\vspace{-10pt}
\end{figure}

As shown in Figure~\ref{fig:wkflw}, a front-end user constructs an approval transaction in six steps.
In {\bf Step 1} and {\bf Step 2}, a front-end user first connects the wallet to a 
{\dapp} and selects desired services provided by the {\dapp}. 
In {\bf Step 3}, the {\dapp} constructs the approval transaction (including sender address, spender address, token address, 
and approval amount) and sends it to the user's wallet for further confirmation. 
After receiving the approval transaction, in {\bf Step 4} and {\bf Step 5}, the user's wallet 
presents the transaction information on its UI and waits for the user to double check
the approval transaction. 
In {\bf Step 6}, once the approval transaction is confirmed by the user, it will be 
forwarded to the blockchain network for validation. Then, the validated transaction
mutates the variable \code{allowance} in the token contract.
It is worth noting that, in {\bf Step 2} and {\bf Step 5} highlighted in
Figure~\ref{fig:wkflw}, the front-end user directly interacts with the UI of the {\dapp} and the wallet to construct approval transactions.

\subsection{The Investigation of {\dapp}s}
\label{subsec:int_dapp}
We collect $22$ {\dapp}s according to the number of received approval transactions and total value being locked (i.e., TVL).  
As of this writing, the TVL of the collected {\dapp}s is over {80\%} and over {50\%} of approval transactions that are related to the selected {\dapp}s.

\noindent \textbf{Investigation Criteria.}\tab
The term \textit{Interpretability} of {\dapp}'s UI indicates the explanatory information
of the approval mechanism. In the investigation of 
{\dapp}'s UI, we evaluate the \textit{Interpretability} based on the following three criterion:
\vspace{-2pt}
\begin{itemize}[leftmargin=10pt]
    \item Does the UI explains the purpose of the approval transaction;
    \item Does the UI presents the existence of the approval transaction;
    \item Does the UI notifies users that approval and execution transactions are consecutively executed.
\end{itemize}
\vspace{-2pt}
As for the \textit{Flexibility}, we check whether the {\dapp}'s UI provides
the modification option for users to customize the approval amount.

\noindent \textbf{Investigation Result.}\tab
Table~\ref{tbl:off_chain_plt} summarizes the investigation result, which reveals that all {\dapp}s
adopt unlimited approval as the default setting on their UIs. Specifically, 
$13$ {\dapp}s do not explain the approval action to users at all, 
and $7$ {\dapp}s partially explain the approval mechanism based on 
our defined criteria. Only \textit{Loopring}~\cite{loopring} and
\textit{Yearn.Finance}~\cite{yearn} comprehensively describe the approval 
mechanism.
As for the \textit{Flexibility}, only 2 {\dapp}s (Bancor~\cite{bancor} and 
Keeper DAO~\cite{keeper}) allow users to adjust their approval amounts on their UIs.

\begin{table*}[t!]
	\caption{\bf The investigation result of {\dapp}s' user interfaces.}
	\label{tbl:off_chain_plt}
	\centering
	\resizebox{0.9\textwidth}{!}{
	\begin{tabular}{lccccc}
		\toprule
		{\bf DApp Name}	      		
		& {\bf TVL (USD)}	
		& {\bf Approval Txn} 
		& {\bf Approval Setting}	
		& {\bf Interpretability} 
		& {\bf Flexibility} \\
		\midrule
		{\it Maker}  	    & $15.305B$			& 0.30\%  	 &Unlimited & -  			& NO\\
		{\it Aave}  	    & $13.084B$			& 1.73\%		 &Unlimited & -  			& NO\\
		{\it Compound}  	& $8.905B$			& 0.43\% 	  	 &Unlimited & -  			& NO\\
		{\it Curve.fi}*  	& $8.419B$			& 1.24\%   	 &Unlimited & -   			& NO\\
		{\it Uniswap}   	& $5.875B$ 	 		& 40.6\%   	 &Unlimited & Criteria 3  	& NO\\
		{\it Sushi Swap}	& $3.488B$	   		& 2.34\% 		 &Unlimited & -  			& NO\\
		{\it Yearn.Finance}*	& $3.331B$ 	   		& 0.33\% 		 &Unlimited & Criteria 1,2,3 & NO\\
		{\it Balancer}		& $1.557B$	   		& 0.69\% 		 &Unlimited & -  			& NO\\
		{\it Bancor}		& $1.378B$	   		& 1.10\%		 &Unlimited & Criteria 1,2  & YES\\
		{\it Alpha Homora}  & $1.042B$          & 0.01\%       &Unlimited & -  			& NO\\
		{\it Cream Finance} & $484.192M$        & 0.19\%       &Unlimited & Criteria 1,2   & NO\\
		{\it Defi Saver}    & $434.392M$        & 0.001\%      &Unlimited & -  			& NO\\
		{\it Keeper DAO}    & $408.48M$         & -       	 &Unlimited & -  			& YES\\
		{\it Harvest Finance} & $399.808M$      & -      	 &Unlimited & -  			& NO\\
		{\it Index Coop}    & $280.553M$        & 0.01\%       &Unlimited & -  			& NO\\
		{\it Set Protocol}  & $198.86M$         & 0.04\%       &Unlimited & -  			& NO\\
		{\it dYdX}          & $179.429M$        & 0.34\%       &Unlimited & Criteria 1,2   & NO\\
		{\it Idle Finance}  & $178.986M$        & 0.013\%      &Unlimited & Criteria 1,2   & NO\\
		{\it Loopring}      & $123.151M$        & 0.05\%       &Unlimited & Criteria 1,2,3 & NO\\
		{\it Integral}      & $90.762M$         & -        	 &Unlimited & -  			& NO\\
		{\it 1inch}         & $33.349M$         & 2.74\%       &Unlimited & Criteria 3   	& NO\\
		{\it 0x}            & $-$              	& 2.29\%       &Unlimited & Criteria 1  	& NO\\
		\bottomrule
 	\end{tabular}}
	\vspace{10pt}
\end{table*}
\begin{figure*}[!htbp]
    \centering
    \begin{subfigure}[b]{.49\textwidth}
        \centering
        \includegraphics[width=1\linewidth]{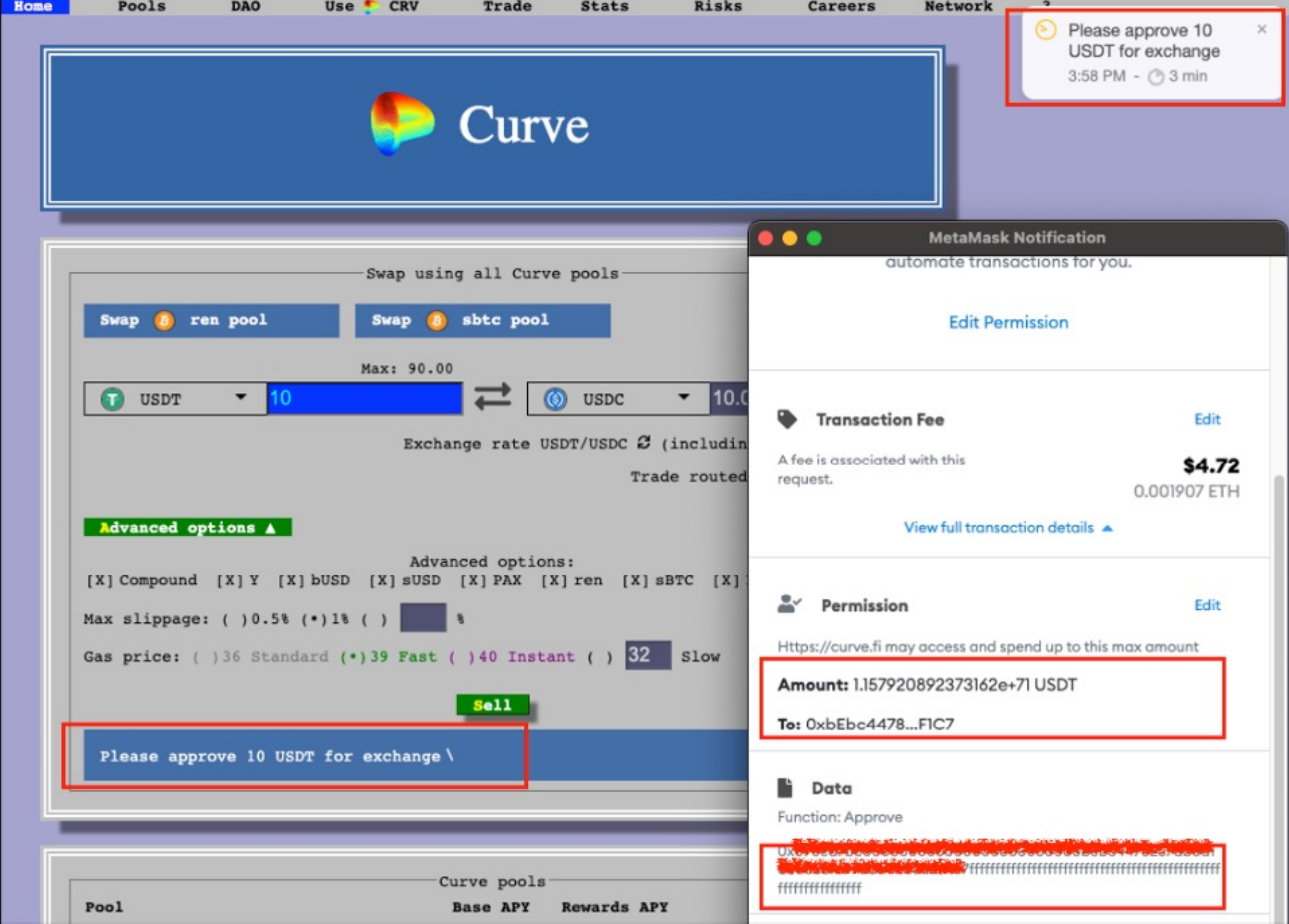} 
        \caption{Curve.fi.} 
        \label{fig:curve}
    \end{subfigure}
    \hfill
    \begin{subfigure}[b]{.49\textwidth}
        \centering
        \includegraphics[width=1\linewidth]{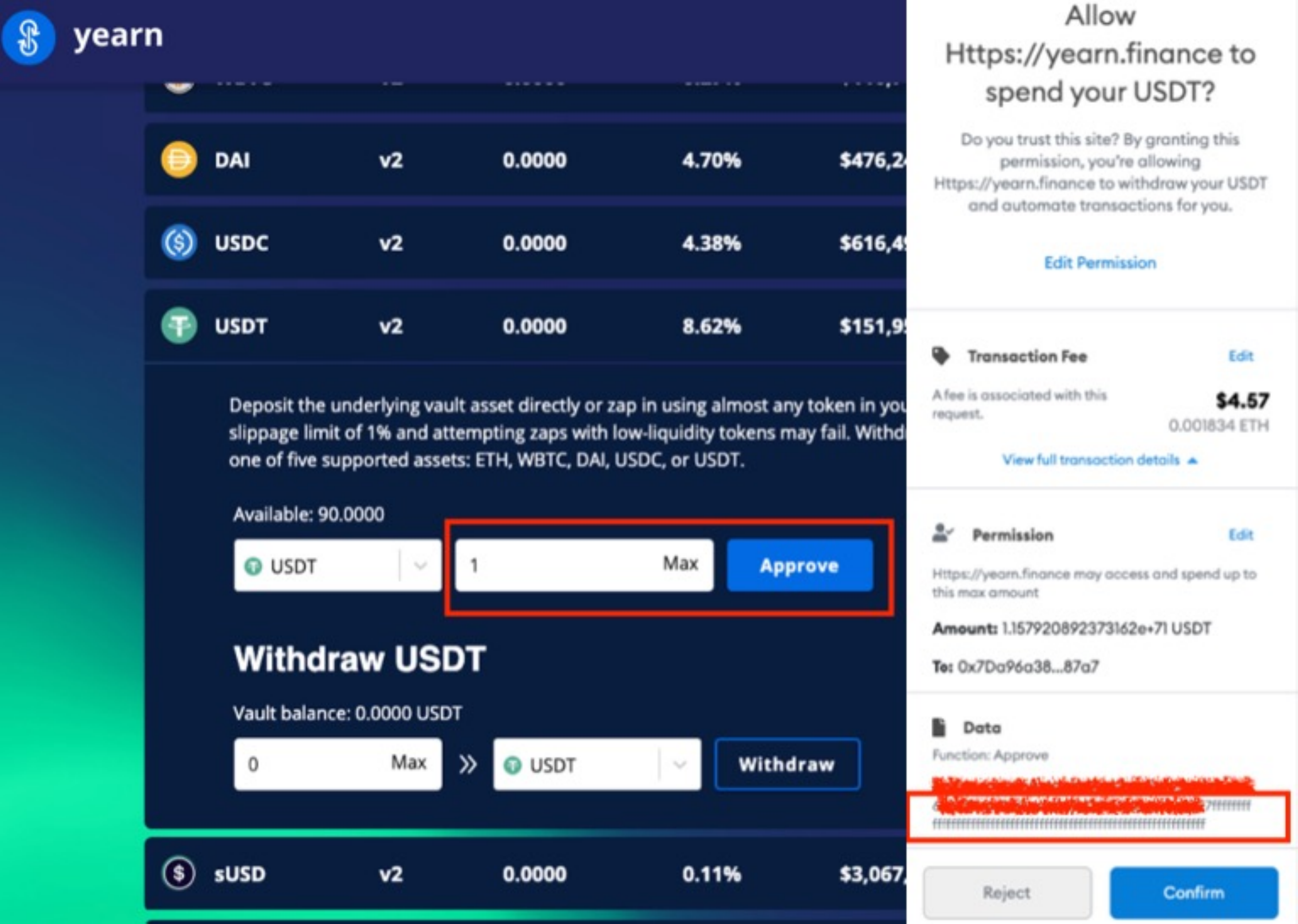} 
        \caption{Yearn.Finance.} 
        \label{fig:yearn}
    \end{subfigure}
    \hfill
    \caption{\bf The demonstration of interacting {\it Curve Finance} and {\it Yearn Finance}.}
    \label{fig:special}
\end{figure*}

\begin{table*}[!t]
    % \vspace{-10pt}
	\caption{\bf The investigation result of wallets' UIs.}
	\label{tbl:off_chain_wallects}
	\centering
	\resizebox{0.9\textwidth}{!}{
	\begin{tabular}{lccccccc}
		\toprule      		
		\multirow{2}{*}{\bf Wallet} 	
		& {\bf Google Play Store} 
		& {\bf Apple App Store}
		& \multicolumn{4}{c}{\bf Interpretability} 
		& \multirow{2}{*}{\bf Flexibility}  \\
		& {\bf Downloadings (\#)} & {\bf Score / Reviews (\#)}& Spender 	& Token & Amount &  Explanation & \\
		\midrule
		{\it Metamask}  	& 10,000,000+ &	4.2 / 7,800+  & \cmark & \cmark & \cmark & \cmark  & YES  \\
		{\it Trust Wallet}  & 10,000,000+ &	4.7 /160,000+ & \xmark & \xmark & \xmark & \xmark  & NO \\
		{\it Crypto.com}  	& 1,000,000+ & 4.3 / 68,000+   & \xmark & \cmark & \cmark & \cmark  & NO \\
		{\it Coinbase}  	& 1,000,000+ & 4.6 / 89,000+	  & \xmark & \xmark & \xmark & \xmark  & NO \\
		{\it Argent}  		& 100,000+ 	 & 4.5 / 1,300+   & \xmark & \cmark & \cmark & \cmark  & YES \\
		{\it Pillar}  		& 100,000+	 & 4.5 / 100+   & \xmark & \xmark & \xmark & \xmark  & NO \\
		{\it MEW}  			& 500,000+ 	 & 4.7 / 4,000+   & \xmark & \xmark & \xmark & \xmark & NO \\
		{\it imToken}  		& 500,000+ 	 & 4.4 /400+   & \cmark & \cmark & \cmark & \cmark  & YES \\
		{\it Rainbow}  		& N/A & 4 / 500+	  & \xmark & \xmark & \xmark & \cmark  & NO \\
		\bottomrule
 	\end{tabular}}

\end{table*}

\noindent \textbf{Special Cases.}\tab
We also discover two special cases, i.e., \textit{Curve.fi} and \textit{Yearn.Finance},
which display misleading information for front-end users to grant unlimited approval. 
Both {\dapp}s request an approval with the amount based on users' spending amount on
their web UI, but, in fact, they prepare an unlimited approval transaction for users.
Figure~\ref{fig:special} shows the process of constructing the
approval transactions for the two cases. 
Specifically, \textit{Curve.fi} is a DEX platform for stable coins, and it is supposed to request an approval with the amount according to users' spending amount, i.e., 10 USDT, as shown in the left bottom and right top of Figure~\ref{fig:curve}.
However, the right bottom of Figure~\ref{fig:curve} suggests that \textit{Curve.fi} directly prepares an unlimited approval transaction without providing any notification related to the unlimited approval on its web UI. 
As a result, users might be misled to approve unlimited tokens to \textit{Curve.fi}. 
Similarly, \textit{Yearn.Finance}
provides misleading information on their web UI as well. In Figure~\ref{fig:yearn}, when a user
believes that she/he only approves 1 \textit{USDT}, \textit{Yearn.Finance} actually provides an unlimited approval transaction by default.

\subsection{The Investigation of Wallets}
\label{subsec:int_wallet}
We collect $9$ well-known wallets based on the number of downloads
ranked by the Google Play Store~\cite{googleplay} 
and the number of reviews in the Apple Store~\cite{applestore}.

\noindent \textbf{Investigation Criteria.}\tab
As aforementioned, the wallet's UI presents the approval transaction
prepared by the {\dapp} to the user for further confirmation. 
To analyze \textit{Interpretability} of wallets' UIs, we mainly investigate the transaction information (i.e., sender (\term{U}), spender (\term{S}) and approval amount) and the explanation of the approval mechanism presented on the wallet's UI.
As for the \textit{Flexibility}, we aim to check whether wallets provide a modification option for users to adjust the approval amount.

\noindent \textbf{Investigation Result.}\tab
Table~\ref{tbl:off_chain_wallects} gives the result.
Specifically, in terms of \textit{Interpretability},
4 (out of 9) wallets do not display any transaction information and
explanation of the approval transaction for users and only \textit{Metamask} 
presents the comprehensive information.
In terms of \textit{Flexibility}, only 33\% wallets (3/9, i.e., 
{\it Metamask}, {\it imToken} and {\it Argent}) embed the modifying
feature on their UIs. This indicates that 67\% (6/9) wallets will
lead users to construct an approval transaction based on {\dapp}s' default setting 
(i.e., unlimited approval). 
Only front-end users of {\it Metamask}, {\it imToken} and {\it Argent} can 
customize their approval amounts to avoid constructing unlimited approval. 
In conclusion, the limited explanation on the approval mechanism and the lack of the 
flexibility on the wallets' UI can aggravate the abuse of unlimited approval.

\begin{framed}
    \noindent \textbf{Answer to RQ 2:} \textit{
    The investigation result suggests that most {\dapp}s and wallets do not provide comprehensive understandings and flexibility for users to mitigate the risk of unlimited approval. 
    Specifically, the result reveals that unlimited approval is adopted by all ($22$) {\dapp}s on their UIs. However, only 10\% {\dapp}s and wallets (i.e., 9\% (2/22) and 11\% (1/9), respectively) provide explanatory information of the approval mechanism.
    Moreover, only 16\% (5/31) of UIs enable users to adjust the approval amounts.
    Surprisingly, we also discover two special {\dapp}s, i.e., {\it Curve.fi} and {\it Yearn.Finance}, which even mislead users to send unlimited approval transactions.}
\end{framed}

\section{User Behavior Analysis }
\label{sec:user}

In this section, we seek to explore the good practice of spending approved tokens without risks. Specifically,
we first detect and sort the user behavior in a temporal order ( Section~\ref{subsec:ub_detect}). 
After that, we characterize the user behavior into five modes and formalize the good practice mitigating the 
risk of approved tokens (Section~\ref{subsec:ub_char}). 
Lastly, we quantify the user behavior based on those modes and the good practice
(Section~\ref{subsec:ub_dist}).

\begin{figure}[h]
	\centering
	\includegraphics[width=1\linewidth]{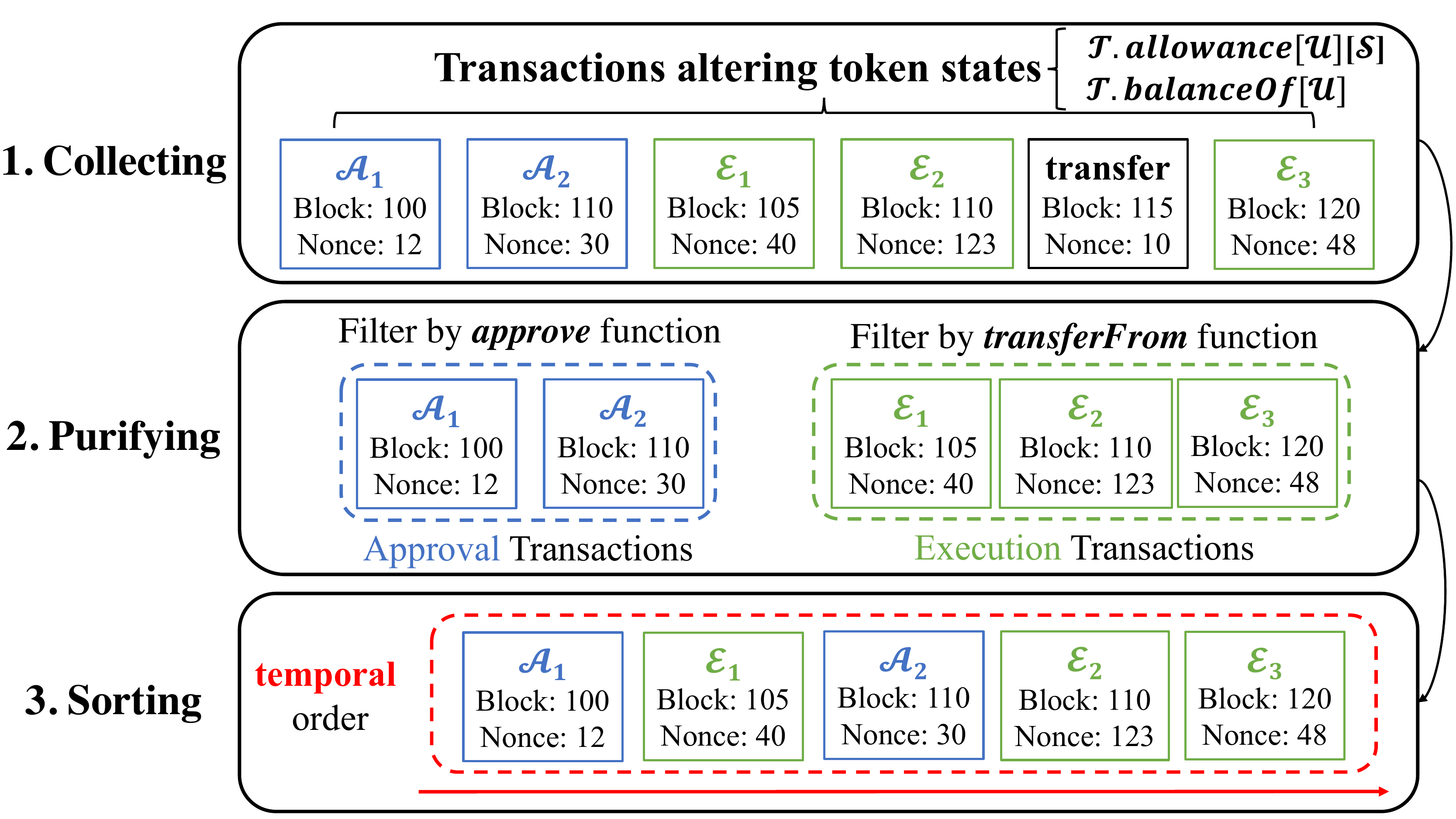} 
	\caption{{\bf The process of user behavior detection with a given tuple (\term{U}, \term{S}, \term{T}).}}
	\label{fig:ubTracing}
\end{figure}
\begin{table*}[ht]
    \caption{\bf The classification of user behaviors.}
	\label{tbl:ubm}
	\centering
    \resizebox{1.0\textwidth}{!}{
    \begin{tabular}{ccll}
		\toprule      		
		Mode ID&{\bf \# of \term{A}, \term{E}} & {\bf Sequence Order} & {\bf Good Practice}   \\
		\midrule
		\multirow{2}{*}{\bf 1}& \multirow{2}{*}{One, One} 
		& \multirow{2}{*}{\term{A}\textsubscript{1} $\rightarrow$ \term{E}\textsubscript{1}}
        & \multirow{2}{*}{\term{OA}\textsubscript{1} $\rightarrow$ \term{E}\textsubscript{1}, where \equa{amount_{\mathcal{OA}_{1}} == amount_{\mathcal{E}_{1}}}}\\
        &&&\\
        \hline
        \multirow{2}{*}{\bf 2}& \multirow{2}{*}{One, Many} 
        & \multirow{2}{*}{\term{A}\textsubscript{1} $\rightarrow$ \term{E}\textsubscript{1} $\rightarrow$ ... $\rightarrow$ \term{E}\textsubscript{n}}
        & \term{OA}\textsubscript{1} $\rightarrow$ \term{E}\textsubscript{1} $\rightarrow$ ... $\rightarrow$ \term{E}\textsubscript{n}, \\
        &&& where n > 1 \&\& \(amount_{\mathcal{OA}_{1}}\) == \(\sum_{i=1}^{n} amount_{\mathcal{E}_{i}}\) \\
        \hline
        \multirow{2}{*}{\bf 3}& \multirow{2}{*}{One or Many, Zero} 
        & \multirow{2}{*}{\term{A}\textsubscript{1} or \term{A}\textsubscript{1} $\rightarrow$ ... $\rightarrow$ \term{A}\textsubscript{n}}
        & \multirow{2}{*}{All cases in this mode are considered as bad practices.}\\
        &&&\\
        \hline
        \multirow{2}{*}{\bf 4}& \multirow{2}{*}{Many, One} 
        & \multirow{2}{*}{\term{A}\textsubscript{1} $\rightarrow$ ... $\rightarrow$ \term{A}\textsubscript{n} $\rightarrow$ \term{E}\textsubscript{1}}
        & \multirow{2}{*}{All cases in this mode are considered as bad practices.} \\
        &&&\\
        \hline
        \multirow{3}{*}{\bf 5}& \multirow{3}{*}{One or Many, One or Many}
        & \multirow{3}{*}{Combination of Mode 1, 2, 3, 4}
        & Consisting of only \( \mathcal{UA} \rightarrow \mathcal{E}_{1} \rightarrow ... \rightarrow \mathcal{E}_{n} \rightarrow \mathcal{ZA} \) or \\
        &&& \( \mathcal{OA} \rightarrow \mathcal{E}_{1} \rightarrow ... \rightarrow \mathcal{E}_{m} \),\\
        &&&where n,m >= 1 \&\& \(amount_{\mathcal{OA}_{1}}\) == \(\sum_{i=1}^{m} amount_{\mathcal{E}_{i}}\)\\
		\bottomrule
    \end{tabular}}
\end{table*}

\subsection{User Behavior Detection}
\label{subsec:ub_detect}

The user behavior indicates a group of ordered approval (\term{A}) and 
execution (\term{E}) transactions. Through analyzing this group of ordered 
transactions, we can understand how a user manages to spend his or her approved 
tokens. In Figure~\ref{fig:ubTracing}, we present the process of detecting
user behaviors for a given tuple (\term{U}, \term{S}, \term{T}). Specifically, there are three steps:
\begin{itemize}[leftmargin=*]
    \item[1)] We collect all transactions altering states of variables: \term{T}.\code{allowance}[\(\mathcal{U}\)][\(\mathcal{S}\)] and \term{T}.\code{balanceOf}[\(\mathcal{U}\)]. However, the extracted transactions also include transactions initiated by \term{U} to directly transfer his or her ERC20 tokens. 
    \item[2)] We purify collected transactions to consist of only the transaction triggered by the function \code{approve} and \code{transferFrom}. Since we only consider the user as an EOA, the \code{approve} and \code{transferFrom} are invoked in separate transactions. 
    \item[3)] We merge purified transactions in the temporal order. It is worth noting that we sort transactions based on their block number and transaction nonce ~\footnote{The transaction nonce implies the order of transactions in the same block.} so that all extracted transactions regarding the user behavior can be arranged in absolute order. 
\end{itemize}

\subsection{User Behavior Characterization}
\label{subsec:ub_char}

Based on the sequence and the amount of \term{A} and \term{E}, we can
characterize user behaviors into five modes, including four basic modes and one compound mode, as listed in Table~\ref{tbl:ubm}. 
\begin{itemize} [leftmargin=*]
    \item \textbf{Mode-1 (aka one-to-one mode): }
    The user only spends approved tokens once after one approval transaction.
    Achieving good practice in mode 1 requires users to
    construct \oa for the \term{A}\textsubscript{1} with the approval amount 
    that equals the number of tokens spending on the following execution 
    transaction \term{E}\textsubscript{1}. Through this, 
    the \term{T}.\code{allowance[\(\mathcal{U}\)][\(\mathcal{S}\)]} will remain zero and 
    the \RL of the users' approved token becomes no-risk. On the other hand, 
    using unlimited approval in mode 1 leads the approved tokens to become risky. 
    \item \textbf{Mode-2 (aka one-to-many mode): }
    The user spends approved tokens several times with only 
    one approval transaction. To achieve the good practice, the user has 
    to build \oa for the \term{A}\textsubscript{1} with the approval amount 
    that equals the total amount of tokens spent on the following 
    execution transactions (\term{E}\textsubscript{1}, ..., \term{E}\textsubscript{n}).
    \item \textbf{Mode-3 (aka only-approval mode): } 
    The user {\bf only} sends the approval transaction(s).
    We consider the user behavior following mode 3 as bad practices because of the 
    redundant approval transactions.
    Even though the user might remain their approved tokens without risk by constructing
    a zero approval transaction at the end like ({\term{A}\textsubscript{1} $\rightarrow$ ...
    $\rightarrow$ \term{A}\textsubscript{n}$\rightarrow$ \term{ZA}}), the user wastes money 
    (gas fee) for only sending approval transactions.
    \item \textbf{Mode-4 (aka many-to-one mode): } 
    The user sends many approval transactions before spending approved tokens.
    The user behavior following mode 4 can be interpreted as the 
    combination of modes 1 and 3, which is a bad practice due to the redundant approval transactions.
    \item \textbf{Mode-5 (aka compound mode): } 
    The user behavior of mode 5 is the combination of at least two 
    basic mode behaviors.
    To achieve good practice in mode 5, the user has to 
    construct approval and execution transactions by only using two fixed patterns: 
    {\it 1) } \( \mathcal{UA} \rightarrow \mathcal{E}_{1} \rightarrow ... \rightarrow \mathcal{E}_{n} \rightarrow \mathcal{ZA} \), 
    {\it 2) } \( \mathcal{OA} \rightarrow \mathcal{E}_{1} \rightarrow ... \rightarrow \mathcal{E}_{m} \) to ensure that the approved token has no risk.
\end{itemize}

\noindent {\bf Achieve the good practice.}\tab
Table~\ref{tbl:ubm} also explains the way to achieve the good practice regarding the \RL of users' approved tokens after the last transaction of the user behavior.
Accordingly, users can mitigate the risk of approved tokens by following the good practice mentioned in modes 1, 2, and 5.
Specifically, the good practice requires \textit{on-demand approval and timely spending}, which means users need to approve the right amount of tokens and trade them as soon as possible based on the approval mechanism, and do not delegate any unnecessary tokens to the {\dapp} thereafter.

\begin{table*}[t]
    \caption{{\bf The user behavior distribution of 10 pairs of \term{S} and \term{T}.} In total, we 
    have detected 1,496,886 user behaviors based on the ten selected pairs.  There are 1,314,995 
    identical users among detected user behaviors. This is because a user might spend approved
    tokens with different pairs.}
	\label{tbl:bt}
	\centering
	\resizebox{1.0\textwidth}{!}{
    \begin{tabular}{ccccccccc}
		\toprule      		
		\multirow{2}{*}{\bf Pair ID} 
        & \multirow{2}{*}{\bf Spender ({\bf \term{S}})} 
        & \multirow{2}{*}{\bf Token ({\bf \term{T}})} 
        & \multirow{2}{*}{\bf Identical User (\# / \%)} 
        & \multicolumn{5}{c} {\bf Behavior Mode Distribution (\%)} \\
        &&&&Mode1&Mode2&Mode3&Mode4&Mode5\\
		\midrule
        {\bf 1}&  UniswapV2& USDT& 338,855 / 6.8\% & 37.1\% & {\bf 41.1\%} & 18.8\% &1.2\% &1.8\% \\
        {\bf 2}&  Compound& USDC& 249,407 / 5.0\%& {\bf 83.1\%} & 7.5\% & 3.0\% &3.5\% &2.8\% \\
        {\bf 3}&  UniswapV2& USDC& 197,474 / 4.0\%& 35.7\% & {\bf 46.6\%} & 14.2\% &1.1\% &2.4\% \\
		{\bf 4}&  Wyern& WETH& 141,242 / 2.8\%& 15.0\% & 13.8\% & {\bf 68.0\%} &0.7\% &2.5\% \\
        {\bf 5}&  UniswapV2& SHIB& 117,381 / 2.4\%& {\bf 38.7\%} & 28.5\% & 30.6\% &1.2\% &1.1\% \\
        {\bf 6}&  UniswapV2& DAI& 115,538 / 2.3\%& 39.3\% & {\bf 42.9\%} & 14.1\% &1.3\% &2.4\% \\
        {\bf 7}&  Shiba& SHIB& 102,162 / 2.1\%& {\bf 52.6\%} & 31.5\% & 6.6\% &4.5\% &4.8\% \\
        {\bf 8}&  UniswapV2& WETH& 81,142 / 1.6\%& {\bf 40.8\%} & 35.5\% & 20.1\% &1.5\% &2.1\% \\
        {\bf 9}&  Wyern& USDT& 78,437 / 1.6\%& {\bf 43.9\%} & 23.2\% & 30.7\% &1.3\% &0.9\% \\
        {\bf 10}& Polygon& Matic& 75,248 / 1.5\%& {\bf 67.3\%} & 26.1\% & 3.9\% &1.3\% &1.3\% \\
        \hline
        \multicolumn{3}{c}{\bf User Behavior Count }& {\bf 1,496,886} & {\bf 687,817(46\%)}& {\bf 451,023(30\%)}& {\bf 297,528(20\%)} & {\bf 26,734(1.7\%)}& {\bf 33,784(2.3\%)}\\
        \bottomrule
    \end{tabular}}
\end{table*}

\begin{figure*}[t]
    \vspace{-10pt}
	\centering
	\includegraphics[width=1\linewidth]{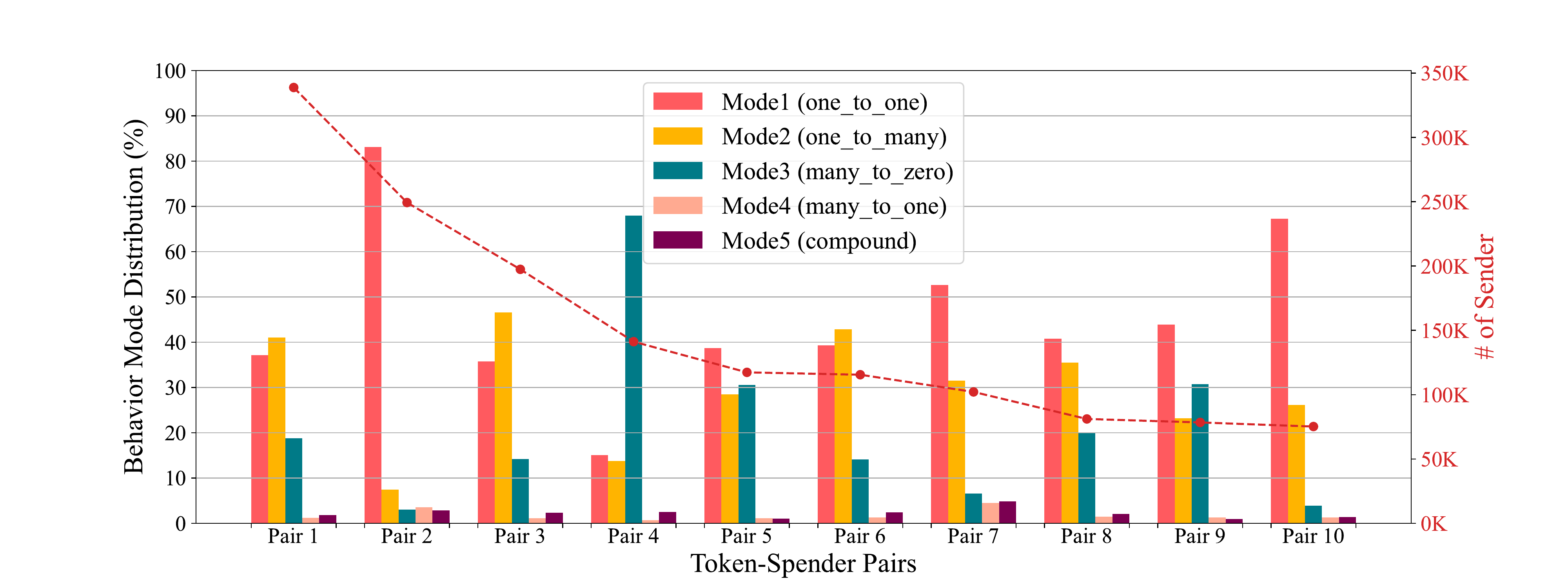} 
    \caption{\bf The distribution of user behaviors. 
    Note that the points on the dot line represent the number of unique users.
    } 
    \label{fig:ub}
\end{figure*}

\subsection{User Behavior Distribution}
\label{subsec:ub_dist}

To conduct a comprehensive analysis of the user behavior, we run the user behavior 
detection for ten representative pairs of \term{S} and \term{T} which cover the most users~\footnote{Of course, \term{S} or \term{T} is of a great amount of TVL or market value as well.}.
As shown in Table~\ref{tbl:bt}, based on these ten pairs, we totally collect {\bf 1,496,886} user behaviors, which covers 1,314,995 (30\% of all collected users in our data set) unique users.
We also measure the distribution of the user behavior based on five defined modes in 
Figure~\ref{fig:ub} and present the corresponding statistics in Table~\ref{tbl:bt}. 
The result shows that the characterized user behaviors are extremely concentrated in 
modes 1, 2 and 3, i.e., 
46\%, 30\% and 20\% of unique users that use approval transactions, respectively.
Specifically, 60\% (6/10) and 30\% (3/10) of pairs are mostly used to spend approved tokens by following modes 1 and 2. 
Meanwhile, in pair 3, 68\% of user behaviors follow mode 3, which only sends approval transactions.
However, only 2.3\% of unique users act based on mode 5 which
can possibly become the good practice.

Moreover, we take a further step to measure the distribution of the user behaviors
following the good practice.
We observe that only 0.01\% (83/1,138,840) of user behaviors follow good practice 
in modes 1 and 2. The reason is that over 99\% of user behaviors belonging to modes 1 and 2 
start with unlimited approval. The fact reveals that most users in mode 1 and mode 2 have 
remained their approved tokens with risk of stealing.
As for the user behavior in mode 5, 7\% (2,392 / 33,784) of user behaviors comply with 
the good practice of using approval transactions. In total, only {\bf 0.2\% (2,475 / 1,496,886)} of
user behaviors follow the good practice to spend approved tokens.
The revealed worrisome result may be caused by {\dapp}'s unclear interpretation and low flexibility regarding the ERC20's approval mechanism.

\begin{framed}
    \noindent \textbf{Answer to RQ 3:} \textit{
    The analysis result suggests that 76\% of user behaviors comply with modes 1 and 2, and 99\% of their user behaviors are using unlimited approval.
    Theoretically, modes 1, 2, and 5 may lead to the good practice. To this end, users shall spend out approved tokens by either granting limited approval on demand or revoking the approved tokens after the last execution transaction.
    However, the result reveals a worrisome fact that only 0.2\% of user behaviors follow the good practice to spend the ERC20 tokens based on the approval mechanism.
    }
\end{framed}

\section{Existing Solutions and Suggestions}
\label{sec:dis}
In this section, we first discuss two existing solutions attempting to address the trade-off between the convenience and security of unlimited approval.
After that, we provide suggestions for the stakeholders (i.e., users, {\dapp}s, and wallets) regarding the approval mechanism to mitigate the risks of unlimited approval.

\subsection{Existing Solutions}
\label{subsec:dis_solu}

\noindent {\bf ERC777.}\tab
Compared to the ERC20 token standard, ERC777~\cite{erc777} has the distinct ability to 
send tokens in a single transaction with the introduction of the \textit{operator}, which is an (arbitrary) address authorized by users
to transfer users' tokens with provided logic (i.e., the hook function). 
Moreover, users have the flexibility to authorize/revoke multiple \textit{operator}s.
Obviously, the advantage of ERC777 is to simplify the process 
of transferring tokens with only one transaction. However, it is also a two-folded 
solution. First, transferring ERC77 tokens requires a high transaction fee due to 
the hook functions. Second, ERC777 also introduces a trust issue that users have to 
select a trustworthy \textit{operator}. Third, spending ERC777 on a new platform requires
users to authorize a new \textit{operator} again.

\noindent {\bf EIP2612.}\tab
EIP2612~\cite{eip2612} introduces a function \code{permit} for ERC20 tokens to 
validate and process users' approval via an off-chain signed message. This proposal can help users save gas costs from sending an approval transaction, as users only need to pay the gas fee for the execution transactions. Moreover, since there is no cost in the approving process, users can customize their approval amounts for their execution transactions. 
Therefore, users will have no risk on their tokens because the variable 
\term{T}.\code{allowance}[\term{U}] will always remain at zero.
However, to the best of our knowledge, only a few tokens 
(e.g., DAI~\cite{dai_712} and UNI~\cite{uniswap_712}) adopt the \code{permit} function in their contracts, which may cause problems for platforms that try to support different tokens in a unified way.
For example, the platform {\it Multichain}~\cite{incident_multichain} was exploited and lost around $44M$ USD. The root cause is that it allows malicious users to invoke the \code{permit} function for ERC20 tokens which do not adopt EIP2612.

\subsection{Suggestions}
In our study, we reveal the fact that unlimited approval is abused in 
the ecosystem. Our findings may shed some light on stakeholders involved in 
the approval mechanism.

\noindent \textbf{ERC20 Token Users.}\tab
Our study reveals that front-end users send at least two transactions (approval and execution 
transactions) to perform actions in a {\dapp}.
Front-end users should be aware of the approval setting designed by the interacting 
{\dapp}s and wallets before approving their tokens. Moreover, users should have security 
consciousness regarding interacted {\dapp}s. For example, for a given {\dapp}, whether the smart contracts are 
verified by some reputable platforms (e.g., Etherscan~\cite{etherscan}), 
or they are officially audited by trustworthy security companies?
To protect approved tokens, we suggest that users only approve the customized amount of tokens needed for 
further executions to minimize the risk. 
Moreover, users should actively monitor their approved tokens so that they can timely revoke approved tokens if necessary.
Users can easily revoke their approved tokens via some platforms (e.g., \textit{approved.zone}~\cite{approvedzone} and\textit{revoke.cash}~\cite{revokecash}).

\noindent \textbf{DeFi {\dapp}s and Crypto Wallets.}\tab
Unlimited approval is often designed as default settings on their web UI to improve the user experience. However, it may mislead novice users without (comprehensive) explanatory information.
As such, {\dapp}s and wallets should precisely explain the risk of unlimited approval so that users can find the suitable approval transaction according to their further execution.  
More importantly, {\dapp}s and wallets also need to enable the feature that allows users to customize their approval amounts, which may further motivate users to construct secure approval transactions. 
Apart from that, it might be a good idea for them to develop revoking functionality on their UIs so that users can timely withdraw their approvals.
Lastly, {\dapp}s should consider adopting emergency operations in their smart contract (e.g., pausing the operation) to protect users' funds.

\section{Related Work}
\label{sec:rel}
\noindent {\bf Analysis for ERC20 Token.}\tab 
In the academic field, several studies have been published to analyze ERC20 tokens~\cite{ chen2019tokenscope, chen2020traveling}.
For example, 
Chen et al.~\cite{chen2019tokenscope} proposed a tool, called \textit{TokenScope}, to investigate inconsistent token behaviors among $7,421$ ERC20 tokens.
Chen et al.~\cite{chen2020traveling} conduct a graph analysis to characterize money transfer,
contract creation, and invocation of ERC20 tokens in the Ethereum ecosystem. 
Besides, some works have addressed the improper use of ERC20 tokens~\cite{rahimian2019resolving, gao2020tracking, xia2021trade}.
For example, Rahimian et al.~\cite{rahimian2019resolving} proposed two protocols to resolve the multiple withdrawal problem raised by the ERC20 token standard.
Gao et al.~\cite{gao2020tracking} revealed the prevalence of counterfeit cryptocurrencies on Ethereum.
Xia et al.~\cite{xia2021trade} systematically analyzed the behavior, the working mechanism, and
the financial impacts of scam tokens and identified over 10K scam tokens and scam liquidity pool on Uniswap V2.

\noindent {\bf DeFi Security.}\tab
Apart from the security issues of the ERC20 token, in recent years, 
DeFi security has been widely researched and has drawn great attention. 
Numerous studies have been conducted to address the security issues like price manipulation~\cite{wu2021defiranger},
front-running~\cite{eskandari2019sok,daian2020flash,zhou2020highfrequency}, 
governance~\cite{gudgeon2020decentralized}, and flash loan attacks~\cite{qin2020attacking}.
Furthermore, Sam et al.~\cite{werner2021sok} systematically studied the security 
issues introduced in theory or occurred in practice.

\noindent {\bf Smart Contract Vulnerability.}\tab
The vulnerability of the smart contract can cause severe loss 
for {\dapp}s and users. 
Many studies dedicated to detecting or protecting vulnerable smart contracts with 
static analysis techniques (e.g., symbolic execution and formal verification)~\cite{luu2016making,kalra2018zeus,tsankov2018securify,grech2018madmax}
or dynamic analysis techniques (e.g., fuzzing)~\cite{jiang2018contractfuzzer,wustholz2020harvey,schneidewind2020ethor, rodler2018sereum}.
Moreover, Perez et al.~\cite{perez2021smart} surveyed 23,327 vulnerable contracts
identified by six academic projects and discovered that only 1.98\% of them have
been exploited.
\vspace{-5pt}
\section{Conclusion}
\label{sec:con}
In this paper, we present the first in-depth study of quantifying the risk of unlimited approval of ERC20 tokens on Ethereum. 
Our study proposes a fully-automatic approach to detect the approval transactions, and 
reveals the prevalence (60\%) of unlimited approval in the ecosystem.
We also conduct an investigation to reveal the security 
issues involved in interacting with 31 UIs (22 {\dapp}s and 9 wallets) to send approval 
transactions. The result shows that only a few UIs provide explanatory understandings (10\%) and flexibility (16\%) for users to mitigate the risk of unlimited 
approval.
Furthermore, we perform the user behavior analysis to characterize five modes of 
user behaviors and formalize the good practice to use approved tokens. The result reveals that
users (0.2\% of user behaviors) barely follow the good practice towards mitigating the 
risks of unlimited approval.
Finally, we discuss two existing solutions attempting to address the
trade-off between convenience and security of unlimited approval, and provide possible suggestions.

\begin{acks}
We sincerely thank our shepherd and the anonymous reviewers for their valuable feedback and suggestions. This work was supported by the National Natural Science Foundation of China (grants No.62172360, No.U21A20467).
\end{acks}

\balance
\bibliographystyle{ACM-Reference-Format}
\bibliography{main}

\newpage

\appendix

\section{The adoption of EIP2612}
\label{apd:EIP2612}

\begin{figure}[b]
    \centering
    \begin{subfigure}[b]{.49\textwidth}
        \label{fig:dai}
        \centering
        \includegraphics[width=1\linewidth]{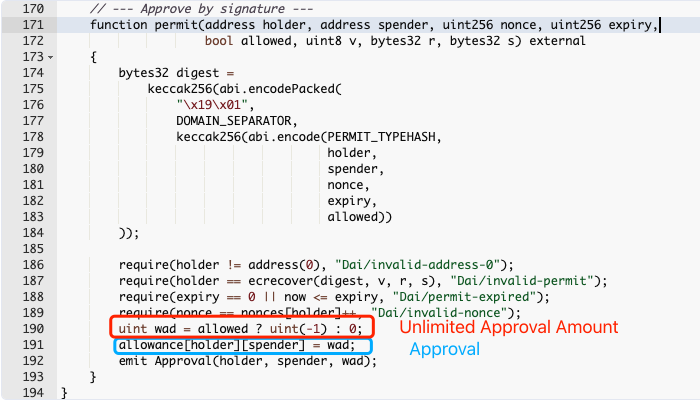}
        \caption{The code of the token {\bf DAI}.} 
    \end{subfigure}
    \hfill
    \begin{subfigure}[b]{.49\textwidth}
        \label{fig:uni}
        \centering
        \includegraphics[width=1\linewidth]{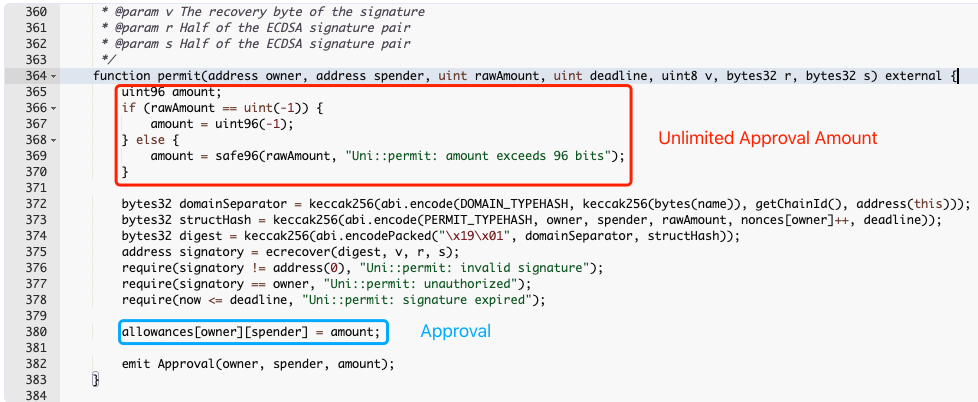} 
        \caption{The code of the token {\bf UNI}.} 
    \end{subfigure}
    \hfill
    \caption{\bf The adoption of EIP2612.}
    \label{fig:permit}
\end{figure}

As aforementioned, with \code{permit} function, users can delegate {\dapp}s to spend their tokens with a customized amount of tokens in one transaction. 
The proposal can help users save gas fees from sending additional approval 
transaction and mitigate the risk of unlimited approval. However, as shown in
Figure~\ref{fig:permit}, DAI and UNI tokens force users to approve unlimited
tokens in the \code{permit} function according to their newest smart contract.

\section{The demonstration of mitigating the risk of unlimited approval}
\label{apd:goodpractic}

\begin{figure*}[!b]
	\centering
	\includegraphics[width=1\linewidth]{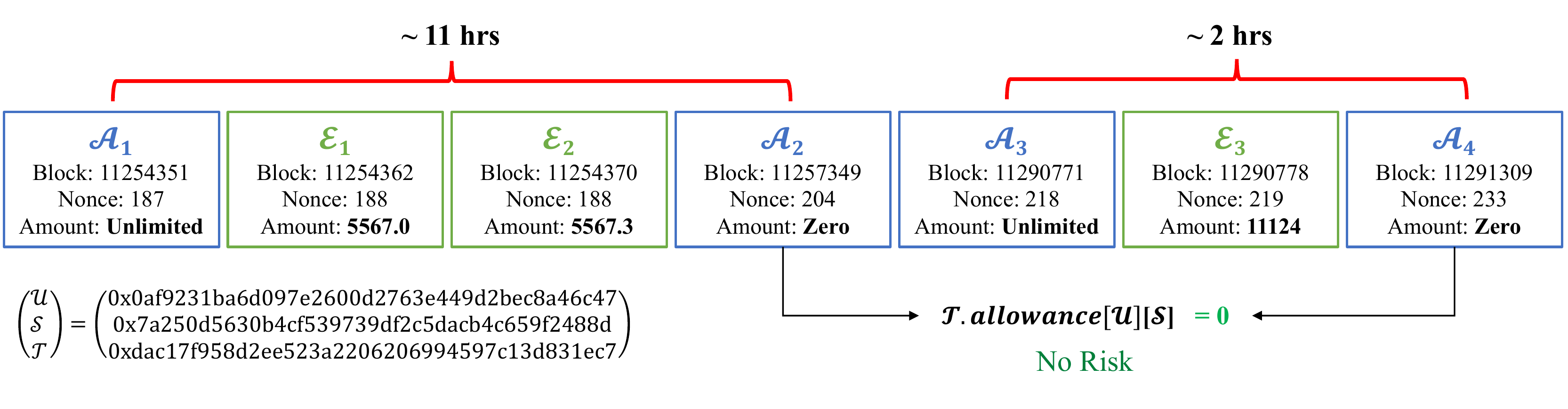}
		\term{S}: Uniswap V2 Router; \term{T}: USDT;
    \caption{\bf An example of mitigating the risk of using unlimited approval.} 
    \label{fig:goodpractice}
\end{figure*}

Figure~\ref{fig:goodpractice} demonstrates a real-world case to present the 
good practice using unlimited approval. 
There are two rounds of spending approved tokens by users. 
The time interval between the two rounds is around 5 days. In the first round,
the user first sends an unlimited approval transaction to delegate his or her 
{\it USDT} to {\it Uniswap V2 Router}. Then, the following two execution transactions
spend {\it 5567} and {\it 5567.3} USDT on {\it Uniswap V2}. Finally, the user sends a zero approval transaction to revoke his or her approved tokens. The whole process takes about 11 hours to complete. The 11 hours is the time difference between the unlimited approval transaction \term{A}\textsubscript{1} and the zero approval transaction \term{A}\textsubscript{2}.
Similarly, it takes about 2 hours for the user to complete his or her second round of spending approved tokens.

% \vspace{-5pt}

\end{document}